\documentclass[prl,twocolumn,english
,aps,reprint,superscriptaddress,notitlepage,nobibnote,preprintnumbers,showpacs]{revtex4-1}

\usepackage{graphicx}
\usepackage{dcolumn}
\usepackage{bm}
\usepackage{epsfig}
\usepackage{gensymb}
\usepackage{mathrsfs}
\usepackage{amsmath}
\usepackage{amssymb}
\usepackage{graphicx,bm}
\usepackage{natbib}
\usepackage{slashed}
\usepackage[makeroom]{cancel}
\usepackage{dsfont}
\usepackage{longtable}
\usepackage{multirow}
 \usepackage[titletoc]{appendix}

\def\fun#1#2{\lower3.6pt\vbox{\baselineskip0pt\lineskip.9pt
  \ialign{$\mathsurround=0pt#1\hfil##\hfil$\crcr#2\crcr\sim\crcr}}}

\usepackage{tikz}
\usepackage{environ}
\makeatletter
\newsavebox{\measure@tikzpicture}
\NewEnviron{scaletikzpicturetowidth}[1]{%
  \def\tikz@width{#1}%
  \def\tikzscale{1}\begin{lrbox}{\measure@tikzpicture}%
  \BODY
  \end{lrbox}%
  \pgfmathparse{#1/\wd\measure@tikzpicture}%
  \edef\tikzscale{\pgfmathresult}%
  \BODY
}
\makeatother
\usetikzlibrary{decorations.pathmorphing,decorations.markings,trees,shapes}

\newcommand{\Tr}{{\rm Tr\hskip 2pt}}
\def\lsim{\mathrel{\rlap{\raise 2.5pt \hbox{$<$}}\lower 2.5pt\hbox{$\sim$}}}
\def\gsim{\mathrel{\rlap{\raise 2.5pt \hbox{$>$}}\lower 2.5pt\hbox{$\sim$}}}

\newcommand{\bea}{\begin{eqnarray}}
\newcommand{\eea}{\end{eqnarray}}

\input epsf

\usepackage{color}

\newcommand{\comment}[1]{}
\def\l{\langle}
\def\r{\rangle}
\newcommand{\mc}[1]{\mathcal{#1}}

\newcommand{\bi}[1]{\textbf{\textit{#1}}}
\newcommand{\vev}[1]{\langle #1 \rangle}
\newcommand{\ket}[1]{| #1 \rangle}
\newcommand{\bra}[1]{\langle #1 |}
\newcommand{\inner}[2]{\langle #1 | #2 \rangle}
\newcommand{\ie}{{\textit{i.e.}}~}

\newcommand{\eq}[1]{\begin{equation}\begin{split} #1 \end{split}\end{equation}}
\newcommand{\eqs}[1]{\begin{align} #1 \end{align}}

\begin{document}

\title{New Selection Rules from Angular Momentum Conservation}

\author{Minyuan Jiang}
\affiliation{
CAS Key Laboratory of Theoretical Physics, Institute of Theoretical Physics,
Chinese Academy of Sciences, Beijing 100190, China.}
\author{Jing Shu}
\affiliation{
CAS Key Laboratory of Theoretical Physics, Institute of Theoretical Physics,
Chinese Academy of Sciences, Beijing 100190, China.}
\affiliation{School of Physical Sciences, University of Chinese Academy of Sciences, Beijing 100049, P. R. China.}
\affiliation{CAS Center for Excellence in Particle Physics, Beijing 100049, China}
\affiliation{Center for High Energy Physics, Peking University, Beijing 100871, China}
\affiliation{School of Fundamental Physics and Mathematical Sciences, Hangzhou Institute for Advanced Study, University of Chinese Academy of Sciences, Hangzhou 310024, China}
\author{Ming-Lei Xiao}
\affiliation{
CAS Key Laboratory of Theoretical Physics, Institute of Theoretical Physics,
Chinese Academy of Sciences, Beijing 100190, China.}
\author{Yu-Hui Zheng}
\affiliation{
CAS Key Laboratory of Theoretical Physics, Institute of Theoretical Physics,
Chinese Academy of Sciences, Beijing 100190, China.}
\affiliation{School of Physical Sciences, University of Chinese Academy of Sciences, Beijing 100049, P. R. China.}

\begin{abstract}\indent
We derive the generalized partial wave expansion for $M \to N$ scattering amplitude in terms of spinor helicity variables. 
The basis amplitudes of the expansion with definite angular momentum $j$ consist of the Poincare Clebsch-Gordan coefficients, while $j$ constrains the UV physics that could generate the corresponding operators at tree level.
Moreover, we obtain a series of selection rules that restrict the anomalous dimension matrix of effective operators and the way how effective operators contribute to some $2 \rightarrow N$ amplitudes at the loop level.

\end{abstract}

\maketitle

\section{Introduction}


Symmetry plays a crucial role in understanding some elegant phenomena of our nature. From Noether's theorem~\citep{Noether:1918zz}, continuous symmetries are always associated with conservation laws. At the quantum level, where the system takes discrete values, some possible transitions of the system from one quantum state to another are forbidden due to symmetry. These phenomena are known as the selection rules, which have been derived for transitions in molecules, atoms, nuclei, or even the elementary particle decay, such as Landau-Yang theorem~\citep{PhysRev.77.242, Landau1948}. 

One crucial task for the high energy physics studies is the precision tests of the Standard Model (SM), for which the SM Effective Field Theory (SMEFT) is a necessary tool. Recently, some interesting progress has been made to apply the on-shell amplitude methods to SMEFT~\citep{Shadmi:2018xan,Ma:2019gtx,Henning:2019enq,Durieux:2019siw,Falkowski:2019zdo}. Moreover, various constraints have been observed on the relations among operators or between operators and observables~\citep{Elias-Miro:2014eia,Jenkins:2013zja,Alonso:2014rga, Bern:2019wie,Cheung:2015aba,Azatov:2016sqh,Craig:2019wmo}, which indicates the existence of some selection rules in particle scattering processes due to conservation laws. 

%
In~\citep{Ma:2019gtx}, we have matched the effective operator basis with on-shell amplitude basis, dubbed operator-amplitude basis correspondence, which helps us write down all dimension 6 operators in the SMEFT without redundancy~\footnote{For more complicated cases, one needs to use the reduced Young Tableau~\cite{Henning:2019enq} or momentum twistors~\cite{Falkowski:2019zdo} to deal with momentum conservation.}. Moreover, we can easily associate the properties of amplitudes to effective operators. 
For $2 \rightarrow2$ scattering, the basis for the partial wave expansion of an amplitude is the Wigner d-matrix $d^j_{\nu\nu'}$, which is precisely an amplitude basis as shown later in the paper. By the operator-amplitude basis correspondence, $d^j_{\nu\nu'}$ induce an operator basis labeled by the angular momentum $j$, which is a recombination of the Warsaw-like basis \citep{Grzadkowski:2010es}. This basis has the privilege that by angular momentum conservation, several selection rules, which restrict the way an operator could contribute to a process, become apparent.

%

%
In this letter, we first define the partial wave basis of multi-particle states and develop the formalism to use the spinor helicity variables to construct the partial wave amplitude basis not only for $2 \to 2$ but also for general $M\to N$ processes. As a byproduct, one can see whether specific processes or operators can be generated at the tree level. Moreover, with this technique, we can obtain two kinds of selection rules for effective operators. Firstly, we find non-trivial constraints on the anomalous dimension matrix that are beyond the non-renormalization relations found in~\citep{Cheung:2015aba}. Then we prove some entirely vanishing contributions from certain operators to specific amplitudes at one loop, part of which is also presented in~\citep{Craig:2019wmo}.

\section{Generalized Partial Waves}


\subsection{Introduction to Spinor-Helicity Variables}

The spinor-helicity variables (or helicity spinors), both for massless and massive particles, are defined as \citep{Arkani-Hamed:2017jhn}
\eq{
	p_{\mu}\sigma^{\mu}_{\alpha\dot\alpha} = p_{\alpha\dot\alpha} = \lambda_{\alpha}^I\tilde{\lambda}_{\dot\alpha I}
,}
where $I$ taking $r=\text{rank}(p_{\alpha\dot\alpha})$ values is called little group index whose contraction reveals the little group invariance of $p$. The hermiticity of $p_{\alpha\dot\alpha}$ requires that $\lambda^*_{\alpha}=\pm\tilde{\lambda}_{\dot{\alpha}}$. 
For massless particles, $r=1$ and $I$ can be omitted, but $\lambda$ and $\tilde\lambda$ are required to have opposite charges under little group $U(1)$, \ie helicities. 
For massive particles, $r=2$ and $I$ is the index for (anti-)fundamental representation of the little group $SU(2)$. It is convenient to set $\det\lambda_\alpha^I=\det\tilde\lambda_{\dot\alpha I}=m$ to satisfy the on-shell condition $p^2=\det(p_{\alpha\dot\alpha})=m^2$.

The helicity spinors constitute the solutions to equation of motion for particles of all spins.
%
%
Therefore all Poincar\'e information -- momentum $p$, spin and its projection $(s,\sigma)$, helicity $h$ -- can be encoded by helicity spinors.
%
Amplitudes can be directly constructed by helicity spinors and should satisfy two conditions: 1. all spinor indices are contracted; 2. little group representations of all external particles should be respected. Hence an all-massless amplitude with helicities $h_i$ (in all-outgoing convention) must take the form
\eq{\label{eq:massless_amplitude}
	\mc{A}_{h_1,\dots,h_N} \sim K(s_{\mc{I}})\prod_i\lambda_i^{r_i}\tilde\lambda_i^{\bar{r}_i},\quad \bar{r}_i-r_i=2h_i,
}
where spinor contractions are omitted. $s_{\mc{I}}=(\sum_{i\in\mc{I}}p_i)^2$ are Mandelstam variables, and the function $K$ contains all the analytic information like poles and branch cuts. The unfactorizable part\footnote{Here ``unfactorizable'' means that the amplitude does not have any kinematic poles or branch cuts on which it could be factorized by unitarity.} is defined as \textit{basis amplitudes} $\mc{B}$ in \citep{Ma:2019gtx}.
When there is a massive particle with spin $s$, the amplitude should have $2s$ totally symmetric little group indices coming from its spinor variables,
\eq{\label{eq:massive_amplitude}
\mathcal{A}^{\{I_1\dots I_{2s}\}}=\lambda^{I_1}_{\alpha_1}\cdots\lambda^{I_{2s}}_{\alpha_{2s}}\mathcal{A}^{\{\alpha_1\dots \alpha_{2s}\}}
}
where $\mathcal{A}^{\{\alpha_1\dots \alpha_{2s}\}}$ also takes the form of eq.~\eqref{eq:massless_amplitude} according to the helicities of the other massless particles, but with $2s$ totally symmetric uncontracted spinor indices.


\subsection{Partial Wave Basis of Multi-Particle States}
\label{sec:partial_wave}

A general multi-particle state is usually written in the tensor representation of Poincar\'e group for a scattering process, which we may call the \textit{tensor basis}.
Now that we can trade the Poincar\'e information $p_i,s_i,\sigma_i$ for the spinor variables $\lambda_i^I, \tilde\lambda_i^J$, we express the tensor basis as
$\ket{\Psi_N}_{\otimes} = \bigotimes_{i=1}^N\ket{\lambda_i^I,\tilde\lambda_i^J,n_i}$.
In this section we are interested in getting the Poincar\'e information for the whole multi-particle state -- the total momentum $P$, total angular momentum $j$ and its projection $\sigma$, hence we need to decompose the tensor basis into an irreducible representation of Poincar\'e group, which we call \textit{partial wave basis}:
%
\eq{\label{eq:basis_partial_wave}
	& \ket{\Psi_N}_j = \ket{P,j,\sigma,a,\{n\}} = \ket{\chi^I,\tilde\chi^J,a,\{n\}},
}
where $\{P,j,\sigma\}$ are replaceable by a pair of auxiliary helicity spinors defined by $P=\chi^I\tilde\chi_I$, $a$ is a label for possible degeneracy, and $\{n\}$ is the collection of particle species information.

Note that in quantum mechanics, we learned about Clebsch-Gordan (CG) coefficients for angular momentum addition, which does not involve the partial wave function, as they are only CG coefficients of the spatial rotation group $SU(2)$. However, here we are talking about the tensor decomposition of Poincar\'e representations. Hence the conversion of basis is described by CG coefficients of the Poincar\'e group instead.
In the following whenever ``CG coefficient'' is mentioned, it means the \textit{Poincar\'e CG coefficient}.
In this letter, we derive the generalized $N$-particle Poincar\'e CG coefficients in terms of helicity spinors as the following overlap function
\eq{
	{}_{\otimes}\inner{\Psi_N}{\Psi_N}_j &= \inner{ \{p_i,s_i,\sigma_i\}^N }{ P,j,\sigma,a } \\
	&\equiv C^{P,j,\sigma,a}_{\{p_i,s_i,\sigma_i\}^N}\delta(P-\sum p_i).
}
We can convert it to a function of helicity spinors by adopting the helicity spinor representation for both state vectors
\eq{
C^{P,j,\sigma,a}_{\{p_i,s_i,\sigma_i\}^N} \equiv \bar{C}^{j,a}_{\{s_i\}^N}(\{\lambda_i,\tilde\lambda_i\}^N,\chi,\tilde\chi).
}
The total angular momentum $j$ is reflected by the requirement that $\bar{C}^{j,a}$ should include $2j$ factors of $\chi$ or $\tilde\chi$ with symmetric little group indices, whose $2j+1$ components give the value of the $\sigma$ label on the left hand side. $\bar{C}^{j,a}$ has the same form of a basis amplitude with an auxiliary particle with spin $j$, which can be expressed similar to eq.~\eqref{eq:massive_amplitude} as
\eq{\label{eq:form_factor}
	\bar{C}^{j,a}_{\{s_i\}^N} = f^{j,a;\{\alpha_1,\dots,\alpha_{2j}\}}_{\{s_i\}^N}\chi^{I_1}_{\alpha_1}\cdots\chi^{I_{2j}}_{\alpha_{2j}},
}
where $f$ is the multi-particle wave function in the all-$\chi$ basis. The wave functions in other basis involving $\tilde\chi$ are equivalent to it via the Dirac equation for $(\chi,\tilde\chi)$ with the mass replaced by $\sqrt{P^2}$.

While in general $f$ can have extra $SU(2)$ indices for massive external particles, we will only be focusing on the multi-massless-particle states that are relevant for massless EFTs, where $s_i$ are replaced by helicities $h_i$.
Let us take the simplest example -- a two massless particle state, with helicity $h_1,h_2$. The CG coefficient in spinor representation $\bar{C}^{j,a}$ is like an amplitude for two massless particles and one massive spin-$j$ particle, whose general form is shown in \citep{Arkani-Hamed:2017jhn} as
\eq{\label{eq:two_massless}
	\bar{C}^j_{h_1,h_2} \sim \frac{[12]^{j+h_1+h_2}}{s^{(3j+h_1+h_2)/2}}(\vev{1\chi}^{j-h_1+h_2}\vev{2\chi}^{j+h_1-h_2})^{\{I_1\dots I_{2j}\}},
}
where we adopt the usual notation that the $\epsilon$ contraction of $\lambda_i\lambda_j$ is denoted by $\vev{ij}$ and that of $\tilde\lambda_i\tilde\lambda_j$ is denoted by $[ij]$. Here the normalization by a power of $s=P^2$ keeps it dimensionless.
The label for degeneracy $a$ is omitted here since we have a unique solution.

Due to angular momentum conservation, the S-matrix is block diagonal in the partial wave basis. Therefore, we can partial wave expand a general scattering amplitude as\footnote{The partial wave expansion for long-range scattering is tricky, which involves zero-poles in other channels and an infinite tower of $j$ in the summation. In this letter, we temporarily ignore such amplitudes, and mainly focus on the basis amplitudes.}
\eqs{\label{eq:amp_j_basis}
	&\mathcal{A}(\{p_i,\sigma_i,n_i\}^N ; \{p'_i,\sigma'_i,n'_i\}^M) \equiv {}_{\otimes}\bra{\Psi_M}\mc{M}\ket{\Psi_N}_{\otimes} \notag\\
	&= \sum_{j,a,b}{}_{j}\bra{\Psi_M}\mc{M}^j_{ab}\ket{\Psi_N}_{j}\sum_{\sigma}C^{P,j,\sigma,b}_{\{p'_i,s'_i,\sigma'_i\}^M}(C^{P,j,\sigma,a}_{\{p_i,s_i,\sigma_i\}^N})^* \notag\\
	&\equiv \sum_{j,a,b}\mathcal{M}^j_{ab}(s)\mc{B}^{j,a\to b}_{\{s_i\}^N \to \{s'_i\}^M}.
}
Since the CG coefficient part $\sum_{\sigma}CC^*$ only involves the Poincar\'e information of the external particles, it is completely determined by symmetry and serves as a basis for a generic amplitude, while the coefficient matrics $\mc{M}^j$ carry the information of the dynamics. We shall call the CG coefficient part with appropriate normalization the \textit{partial wave amplitude basis} $\mc{B}^j$, which is nothing but a special choice of amplitude basis that have definite angular momenta. The sum over $\sigma$ could be translated to the contraction of little group indices in helicity spinor representation
\eq{\label{eq:amp_basis_C}
	& \mc{B}^{j,a\to b}_{\{s_i\}^N \to \{s'_i\}^M} = \sum_{\sigma}C^{P,j,\sigma,b}_{\{p'_i,s'_i,\sigma'_i\}^M}(C^{P,j,\sigma,a}_{\{p_i,s_i,\sigma_i\}^N})^* \\
	&\quad = (\bar{C}^{j,b}_{\{s'_i\}^M})^{I_1,\dots I_{2j}}(\bar{C}^{j,a}_{\{s_i\}^N})^*_{I_1,\dots I_{2j}}.
}
If we write them in the all-$\ket{\chi}$ basis, we can further simplify it using the identity $\chi^I_{\alpha}\chi^J_{\beta}\epsilon_{IJ} = -\sqrt{s}\epsilon_{\alpha\beta}$ and get
\eq{\label{eq:amp_basis_f}
	\mc{B}^{j,a\to b}_{\{s_i\}^N \to \{s'_i\}^M} = (-\sqrt{s})^{2j}f^{j,b}_{\{s_i\}^N}(f^{j,a}_{\{s'_i\}^M})^* .
}

Taking complex conjugation of wavefunctions is equivalent to flipping all the helicities, for instance $(f^j_{h_1,h_2})^* = (-1)^{3j+h_1-h_2}f^j_{-h_1,-h_2}$. Thus we can rederive the Wigner d-matrix as the partial wave amplitude basis of $2\to 2$ scattering by plugging eq.~\eqref{eq:two_massless} into eq.~\eqref{eq:amp_basis_f}
\eq{\label{eq:partial_wave_2to2}
	&\mc{B}^j_{\{h_1,h_2\}\to\{h_3,h_4\}} \sim \frac{(-1)^{j-\Delta}}{s^{2j+h/2}}[12]^{j-h_1-h_2}[34]^{j+h_3+h_4} \\
	&\qquad\times \sum_{i}w_i\vev{13}^{i}\vev{24}^{\Delta-\Delta'+i}\vev{14}^{j-\Delta-i}\vev{23}^{j+\Delta'-i}, \\
	&w_i = \frac{(2j)!(j+\Delta)!(j-\Delta)!(j+\Delta')!(j-\Delta')!}{(j+\Delta-i)!(\Delta'-\Delta+i)!i!(j-\Delta'-i)!}.
}
where we defined $h=-h_1-h_2+h_3+h_4$ as the total helicities in all-outgoing convention, $\Delta = h_2-h_1$, $\Delta' = h_4-h_3$. To understand what this result means, we go to the center of mass frame where we have $[12]=-[34]=\sqrt{s}$, $-\vev{13}=\vev{24}=\sqrt{s}\sin\frac{\theta}{2}$, $\vev{14}=\vev{23}=\sqrt{s}\cos\frac{\theta}{2}$, $\theta$ being the scattering angle. With these substitutions, we recover the Wigner d-matrix
\eq{
	&\mc{B}^j_{\{h_1,h_2\}\to\{h_3,h_4\}} \sim \sum_i(-1)^{-\Delta+h_3+h_4-i}w_i \\
	&\quad\times\left[\cos\frac\theta2\right]^{2j-\Delta+\Delta'-2i}\left[\sin\frac\theta2\right]^{\Delta-\Delta'+2i} \sim d^j_{\Delta,\Delta'}(\theta),
}
where omitted overall factors could be absorbed into the normalizations of the CG coefficients.


\subsection{Single-Particle Resonance and Bridge-Counting Method}

%
One immediate application of partial wave expansion is to determine the spin of resonance in a particular channel.
A basis amplitude generated by an effective operator comes from integrating out massive degrees of freedom at high energies.
Such features are captured by poles and branch cuts in the coefficient function $\mc{M}^j_{ab}(s)$ defined in eq.~\eqref{eq:amp_j_basis}, while poles indicate single-particle resonances and branch cuts indicate multi-particle resonances.
Diagrammatically, they represent the two ways to obtain an effective operator -- tree-level generation and loop-level generation, the latter suppressed by loop factors. Therefore it is phenomenologically interesting to classify operators by the way they are generated.

If we can determine the angular momentum of a basis amplitude, we immediately fix the spin of possible single-particle resonance, which characterizes the UV physics of tree-level generation.
For instance, if $j\geq2$ in a channel, the basis amplitude could not be generated at tree level by a single elementary resonance in the same channel.
Furthermore, by the assumption of tree-level generation, we can use the angular momentum to obtain the type of couplings needed in the UV physics. As an example, the 4 point basis amplitude $\mc{B}=\vev{12}\vev{13}$, which is generated by an operator $F_{\mu\nu}\psi\sigma^{\mu\nu}\psi\phi$, has $j=1$ in $\{1,4\}\to\{2,3\}$ channel which demands the couplings $VF\phi$ and $V\psi\psi$ where $V$ is the vector resonance. The other topology of the tree diagram requires a $j=1/2$ resonance in the $\{1,2\}\to\{3,4\}$ channel, which demands the couplings $F\psi\Psi$. Due to the non-existence of these couplings in renormalizable UV theories \citep{Arzt:1994gp,deBlas:2017xtg,Craig:2019wmo}, we can exclude the tree level generation of $\mc{B}$ and the corresponding operator.
To classify operators more efficiently, instead of constructing the partial wave amplitude basis as in the last section, we wish to compute the angular momentum of a given basis amplitude. It seems easy by observing the fact that the form of eq.~\eqref{eq:amp_basis_f} exhibits $2j$ spinor contractions between the initial state wave function and final state wave function, both consisting of spinor helicity variables.
We may call the spinor brackets contracting spinors from both sides of a scattering channel the \textit{bridges}, and the number of them in a basis amplitude indicates the total angular momentum
\eq{
	j = \frac12\#{\rm (bridges)}.
}
This \textit{bridge-counting method} is simple and efficient for amplitudes that involve few particles and momenta.
However, two preconditions for the counting should be kept in mind: 1. the two wave functions in eq.~\eqref{eq:amp_basis_f} should be both in all-$\chi$ basis, hence the bridges should be all $\vev{\cdot}$ (other forms can always be converted to this form); 2. the bridges should be symmetrized over particles on both sides of the channel.
As a quick example, in the $\{1,2\}\to\{3,4\}$ channel, amplitude $\mc{B}_+=\vev{13}\vev{24}+\vev{14}\vev{23}$ has 2 bridges and thus $j=1$, but $\mc{B}_-=\vev{13}\vev{24}-\vev{14}\vev{23}$ shouldn't be counted as 2 bridges because it's not symmetrized over the bridge contractions. Actually, the latter equals $\vev{12}\vev{34}$ by Schouten identity and has 0 bridges and $j=0$.


\subsection{Poincar\'e Algebra in Helicity Spinor Representation}

While the bridge-counting method is sometimes convenient, it often becomes cumbersome when it comes to more complex amplitudes, which one needs to convert to the form that satisfies the two preconditions above. In this section, we propose an alternative method to obtain angular momentum and more.

In quantum mechanics, we use the non-relativistic $J^2$ operator to obtain angular momentum of, say, a wave function. But in relativistic scenario, we need to use the Pauli-Lubanski operator $W_\mu = \frac12\epsilon_{\mu\nu\rho\sigma}P^{\nu}M^{\rho\sigma}$, which induces a Casimir invariant $W^2$ for Poincar\'e group with eigenvalue $-P^2j(j+1)$, where $j$ is the covariant version of total angular momentum. While this operator is usually applied to single-particle states to classify free particles, it can also be applied to multi-particle states.

Here we propose to use the helicity spinor representation so that the operator can act on the form factors $f$ in eq.~\eqref{eq:form_factor} or basis amplitudes $\mc{B}$. The conformal algebra in the helicity spinor representation was already given in~\citep{Witten:2003nn}, where\footnote{We are using a slightly different normalization than that in~\citep{Witten:2003nn}.}
\eq{\label{eq:Lorentz_alg_spinor_rep}
	& M_{\mc{I},\alpha\beta} = i\sum_{i\in\mc{I}} \big(\lambda_{i\alpha}\frac{\partial}{\partial\lambda_i^{\beta}} + \lambda_{i\beta}\frac{\partial}{\partial\lambda_i^{\alpha}}\big), \\
	& \tilde{M}_{\mc{I},\dot\alpha\dot\beta} = i\sum_{i\in\mc{I}} \big(\tilde\lambda_{i\dot\alpha}\frac{\partial}{\partial\tilde\lambda_i^{\dot\beta}} + \tilde\lambda_{i\dot\beta}\frac{\partial}{\partial\tilde\lambda_i^{\dot\alpha}}\big).
}
$M$ and $\tilde{M}$ are induced from the Lorentz generator as $ M_{\mu\nu}\sigma^{\mu}_{\alpha\dot\alpha}\sigma^{\nu}_{\beta\dot\beta} = \epsilon_{\alpha\beta}\tilde{M}_{\dot\alpha\dot\beta} + M_{\alpha\beta}\bar{\epsilon}_{\dot\alpha\dot\beta}$. The sum is taken over particles in a group $\mc{I}$ for which we want to compute angular momentum, hence for an amplitude we sum over only the initial or only the final state particles. It defines a scattering channel $\mc{I}\to\bar{\mc{I}}$ that the angular momentum is associated with ($\bar{\mc{I}}$ is the complement of $\mc{I}$).

From eq.~\eqref{eq:Lorentz_alg_spinor_rep}, we note the following properties (similar for $\tilde{M}$)
\eq{\label{eq:M_property}
	&M_{\mc{I}}\vev{ij} = 0 \quad {\rm if}\ i,j\in\mc{I}\ {\rm or}\ i,j\in\bar{\mc{I}}, \\
	&M_{\mc{I}}\vev{ij} = i (\ket{i}\bra{j} + \ket{j}\bra{i}),\quad {\rm if}\ i\in\mc{I},j\in\bar{\mc{I}},
}
which inspiringly show that only bridges contribute!
Using eq.~\eqref{eq:Lorentz_alg_spinor_rep} and $P_{\mc{I}}=\sum_{i\in\mc{I}}\lambda_i\tilde\lambda_i$, the Casimir invariant takes the following form
\eq{\label{eq:Casimir_W2}
	W_{\mc{I}}^2(\mc{B}) = &\frac{P_{\mc{I}}^2}{8}\Big(\Tr\tilde{M}^2_{\mathcal{I}}(\mc{B})+\Tr M^2_{\mc{I}}(\mc{B})\Big) \\
&-\frac{1}{4}\Tr(P^{\mathsf{T}}_{\mc{I}} M_{\mathcal{I}}(\mc{B})P_{\mc{I}}\tilde{M}_{\mathcal{I}}(\mc{B}))
}
where $M^2_{\mc{I},\alpha\beta}(\mc{B})\equiv M_{\mc{I},\alpha}^{\quad\gamma}M_{\mc{I},\gamma\beta}\mc{B}$. 
It is tempting to show the conservation of angular momentum defined by this operator. Recall the properties of eq.~\eqref{eq:M_property}, we can easily prove
\eq{
	M_{\mathcal{I}}(\mc{B})=-M_{\bar{\mathcal{I}}}(\mc{B}),\quad \Tr M^2_{\mathcal{I}}(\mc{B})=\Tr M^2_{\bar{\mathcal{I}}}(\mc{B}).
}
Together with $P_{\mc{I}} = -P_{\bar{\mc{I}}}$, we find $W^2_{\mc{I}}\mc{B} = W^2_{\bar{\mc{I}}}\mc{B}$, which means that for any channel $\mc{I}\to\bar{\mc{I}}$ of an amplitude $\mc{B}$, the angular momentum defined by the operator $W_{\mc{I}}^2$ is the same for both initial and final states.

We also show that the operator $W_{\mc{I}}^2$ has the correct eigenvalues. Let's take a simplest example $\mc{B} = \vev{13}$, an amplitude generated by a dimension 5 operator $\mc{O}=\psi_1\phi_2\psi_3\phi_4$. In the channel $\{1,2\}\to\{3,4\}$ where the angular distribution is $d^{1/2}_{\frac12,-\frac12}$, we can read out the angular momentum $j=1/2$ to be checked. Since the amplitude only consists of $\lambda$s, we have $\tilde{M}\vev{13}=0$ thus only the second term in eq.~\eqref{eq:Casimir_W2} survives
\eq{
	(M_{\{1,2\}}^2)_{\alpha}^{\ \beta}\vev{13} &= i M_{\{1,2\},\alpha}^{\qquad\,\gamma}\left(\ket{1}_{\gamma}\bra{3}^{\beta} + \ket{3}_{\gamma}\bra{1}^{\beta}\right) \\
	&= 4\ket{1}_{\alpha}\bra{3}^{\beta} - \vev{13}\delta_{\alpha}^{\beta},
}
hence
\eq{
	W_{\{1,2\}}^2\vev{13} = \frac{(p_1+p_2)^2}{8}\left[4\vev{31} - 2\vev{13}\right] = -\frac{3}{4}s\vev{13}.
}
The eigenvalue $-P^2_{\mc{I}}j(j+1) = -\frac34s$ confirms that $j=1/2$.

In general, we get the partial wave amplitude basis by diagonalizing the $W^2$ representation matrix in the space of basis amplitudes with the same helicity states and the same dimension. The procedure is described in more detail in the appendix.

\section{Selection Rules}

In \citep{Ma:2019gtx,Shadmi:2018xan} it is proposed that operators subject to equation of motion (EOM) and integration by part (IBP) should one-to-one correspond to the unfactorizable amplitude it generates, dubbed ``operator-amplitude basis correspondence''. With appropriate choice of amplitude basis, we could easily translate selection rules for amplitudes to those for operators.
In particular, the operator corresponding to the partial wave basis amplitude $\mc{B}^{j,a\to b}$ should only have matrix elements between the partial wave basis states $\ket{j,\sigma,a}$ and $\ket{j,\sigma',b}$, proportional to $\delta_{\sigma\sigma'}$.
It means that the operator, which we may denote as $\mc{O}^{j,ab}$, would annihilate any states other than the two partial wave basis up to crossing symmetry. Therefore, in the channel that $\mc{B}^{j,a\to b}$ is defined, the operator acting on the tensor basis state picks out only the particular partial wave states $(j,a)$ or $(j,b)$. Unlike the degeneracy $a,b$, the angular momentum $j$ is conserved throughout the whole physical scattering. With this property, we propose the selection rules in two types of calculations, operator renormalizations and loop amplitudes, from angular momentum conservation.


\subsection{Renormalization}

From the on-shell perspective, an effective operator $O_m$ is renormalized by another operator $O_n$ when the loop amplitude with $O_n$ insertion contains UV divergence proportional to the basis amplitude corresponding to $O_m$, characterized by the coefficient $\gamma_{mn}$ as follows
\eq{\label{eq:renorm}
	16\pi^2\mathcal{A}^{\text{1-loop}}_{\text{UV}} =-(\sum_{m,n}\gamma_{mn}C_n\mc{B}_m + \mc{A}')\frac{1}{\epsilon},
}
where $\mc{A}'$ is a factorizable term that may show up in certain diagrams.

\tikzset{
    photon/.style={decorate, decoration={snake,segment length=2.5mm}},
    electron/.style={draw=blue, postaction={decorate},
        decoration={markings,mark=at position .55 with {\arrow[draw=blue]{>}}}},
    gluon/.style={decorate, draw=magenta,
        decoration={coil,amplitude=4pt, segment length=5pt}}
}

\begin{figure}
\centering
\begin{scaletikzpicturetowidth}{0.5\textwidth}
\begin{tikzpicture}[scale=\tikzscale]
\node[rectangle,fill]  (amp1) at (0,0) {};
  \node at (-0.5,0) {$O_n$};
    \node at (-1.4,0) {$\{n_i\}$};
\fill (-0.9,0.3) circle (2pt);
\fill (-1,0) circle (2pt);
\fill (-0.9,-0.3) circle (2pt);
\fill (2.9,0.3) circle (2pt);
\fill (3,0) circle (2pt);
\fill (2.9,-0.3) circle (2pt);
\draw[->] (3.5,0)--(5.5,0);
  \node at (4.5,0.2) {UV divergence};
\node[draw,circle,fill=gray,inner sep=12pt] (amp2) at (2,0) {$$};

\draw[thick] (amp1) to [out=45,in=180] (1,0.7) node[label=$$] {};
\draw[thick] (amp1) to [out=-45,in=180] (1,-0.7) node[label=$$] {};
\draw[thick] (amp2) to [out=135,in=0] (1,0.7) node[label=$$] {};
\draw[thick] (amp2) to [out=-135,in=0] (1,-0.7) node[label=$$] {};
\draw[thick] (amp1) to [out=-135,in=45] (-1,-1) node[label=$$] {};
\draw[thick] (amp1) to [out=135,in=-45] (-1,1) node[label=$$] {};
\draw[thick] (amp2) to [out=45,in=-135] (3,1) node[label=$$] {};
\draw[thick] (amp2) to [out=-45,in=135] (3,-1) node[label=$$] {};
  \begin{scope}[xshift=7.5cm]
 \node[rectangle,fill]  (amp1) at (0,0) {};
  \node at (-0.5,0) {$O_m$};
\node at (-1.4,0) {$\{n_i\}$};
\fill (-0.9,0.3) circle (2pt);
\fill (-1,0) circle (2pt);
\fill (-0.9,-0.3) circle (2pt);
\fill (0.9,0.3) circle (2pt);
\fill (1,0) circle (2pt);
\fill (0.9,-0.3) circle (2pt);

\draw[thick] (amp1) to [out=45,in=-135] (1,1) node[label=$$] {};
\draw[thick] (amp1) to [out=-45,in=-45] (1,-1) node[label=$$] {};
\draw[thick] (amp1) to [out=-135,in=45] (-1,-1) node[label=$$] {};
\draw[thick] (amp1) to [out=135,in=-45] (-1,1) node[label=$$] {};
  \end{scope}
\end{tikzpicture}
\end{scaletikzpicturetowidth}
\caption{Renormalization of an effective operator $O_{m}$ by $O_{n}$ at one loop when $\mc{A}'=0$. $\{n_i\}$ is a collection of external legs shared by both $O_n$ and $O_m$.} \label{fig:RG}
\end{figure}
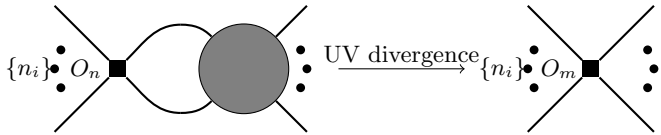

Suppose that there are at least two external legs in a diagram for $\mc{A}^{\rm 1-loop}$ that are shared by both $O_n$ and $O_m$, as shown in fig.~\ref{fig:RG}, the angular momenta $j$ as well as the degenerate label $a$ should both be the same in the two basis amplitudes $\mc{B}_n$ and $\mc{B}_m$. If not, this diagram should have vanishing contribution to $\gamma_{mn}$, up to subtleties caused by $\mc{A}'$. We will have an example to explain the subtlety later, but let us focus on the selection rule assuming $\mc{A}'=0$.

This non-renormalization phenomenon is independent of the criteria proposed in~\cite{Cheung:2015aba},
in which operators are classified by the (anti-)holomorphic weights $(w_i,\bar{w}_i)$ and are renormalized only by the operators of lower weights.
Notably, our selection rule further picks out zero entries in the block of the anomalous dimension matrix allowed by~\cite{Cheung:2015aba}.
We also find cases where a whole type of operators can not be renormalized by another type with lower (anti-)holomorphic weights.
These will be more clear in the following discussion, where we apply our selection rule to cases at dimension 6 and 8.

\bm{$d=6$} To understand the patterns of renormalization at dimension 6 from angular momentum, we list all types of dimension 6 operators\footnote{We use $(\phi, \psi_\alpha, \bar{\psi}_{\dot \alpha}, F_{\alpha \beta},\bar F_{\dot \alpha \dot \beta})$ to denote generic fields transforming under Lorentz group $SU(2)_L \times SU(2)_R \equiv SO(3,1)$ as (0,0), (1/2,0), (0,1/2), (1,0) and (0,1). And $D$ is the covariant derivative. We also use $(\phi,\psi^\pm,F^\pm)$ to denote on-shell scalar, fermion, and vector particles with $\pm$ helicities in scattering amplitudes.} (except for $F^3$ and suppressing possible flavor and Lorentz structures) in table~\ref{tab:d6RG}.
The ``channels'' means the group of two particles shared by $O_m$ and $O_n$ for which we find the angular momentum $j$ using the technique developed previously.
%
We list all dimension 6 operators that could generate the two particles in the row and place them in the column according to $j$.
%
With this arrangement, a diagram specified by the shared particles exists for the renormalization between any two operators in a row, but only operators appearing in the same entry could renormalize each other due to the selection rule.

This table is special as a classification of operators since they can appear multiple times in it.
First, for a type of operators, there may be terms with different Lorentz structures that have different angular momentum in the same channel. Hence the type may appear in multiple columnes in a row, like $\psi^4$ in Table~\ref{tab:d6RG}, whose $j=0$ and $j=1$ basis could renormalize $\psi^2\phi^3$ and $F\psi^2\phi$ respectively but not interchangeable. In SMEFT, it asserts that among $O_{lequ}^1$ and $O_{lequ}^3$ (the definitions are given in appendix~\ref{sec:d6ops}), $O_{eH}$ can only be renormalized by the former while $O_{eW}$ only by the latter \cite{Jenkins:2013zja,Jenkins:2013wua,Alonso:2013hga}.
Second, even for a single operator, it appears in several rows for the different channels we can examine.
%
Therefore, two operators not appearing in the same entry at one row may be in the same entry at another row, which is why we say the selection rule is at the diagram level.

\begin{table}
    \centering
    \begin{tabular}{c|p{1.7cm}<{\centering} p{1.7cm}<{\centering} p{1.7cm}<{\centering}}
        Channels & $j=0$ & $j=1/2$ & $j=1$\\
        \hline
        $F^+F^+$ & $F^2\phi^2(2,6)$ & & \\
        $F^+\psi^+$ & & $F\psi^2\phi(2,6)$ & \\
        $F^+\phi$ & & & $F\psi^2\phi(2,6)$, $F^2\phi^2(2,6)$ \\
        $\psi^+\psi^+$ & $\psi^4(2,6)$, $\psi^2\bar{\psi}^2(4,4)$, $\psi^2\phi^3(4,6)$ & & $\psi^4(2,6)$, $F\psi^2\phi(2,6)$ \\
        $\psi^+\psi^-$ & & & $\psi\bar{\psi}\phi^2D(4,4)$ \\
        $\psi^+\phi$ & & $\psi^2\phi^3(4,6)$, $F\psi^2\phi(2,6)$, $\psi\bar{\psi}\phi^2D(4,4)$ & \\
        $\phi\phi$ & $\phi^4D^2(4,4)$, $\psi^2\phi^3(4,6)$, $\phi^6(6,6)$ & & $\psi\bar{\psi}\phi^2D(4,4)$, $\phi^4D^2(4,4)$
    \end{tabular}
    \caption{Dimension 6 operators classified by their angular momentum in the specified channel. Numbers in the bracket are the (anti-)holomorphic weights $(w,\bar{w})$, so that one can further obtain non-renormalization relations for operators in the same entry by~\citep{Cheung:2015aba}.}
    \label{tab:d6RG}
\end{table}

%

%

Our $j$ selection rule sometimes mixes with the selection rule of gauge charges, like isospin $I$ of $SU(2)_L$. One example in SMEFT is the $H^4D^2$ with both $j=0,1$ and $I=0,1$ in a $\{H^\dagger,H\}$ channel. We denote the couplings in the amplitude $\mathcal{A}(H_1,H^{\dagger}_2,H_3,H^{\dagger}_4)$ with definite $(j,I)$ quantum numbers in $\{1,2\}\to\{3,4\}$ channel as $C^{j,I}$. Due to spin statistics, there are only two independent operators, namely $O_{HD}$ and $O_{H\Box}$ in Warsaw basis \citep{Grzadkowski:2010es}. In terms their Wilson coefficients, we derive
\eq{
	C^{0,0}=3C_{H\Box},	\hspace{1.7cm}	& C^{0,1}=C_{HD}-C_{H\Box}, \\
	C^{1,0}=-C_{H\Box}-C_{HD},\quad	& C^{1,1}=-C_{H\Box}.
}
From Table~\ref{tab:d6RG}, the $\psi\bar{\psi}^2\phi^2D$ type operators are only renormalized at $j=1$, while in SMEFT we have $O^1_{Hl},O_{He},O^1_{Hq},O_{Hu},O_{Hd}$ of this type with $I=0$ that are only renormalized by the combination $C^{1,0}$, and also $O^3_{Hl}, O^3_{Hq}$ of this type with $I=1$ that are only renormalized by the combination $C^{1,1}$. These are verified by the result in~\cite{Jenkins:2013zja,Jenkins:2013wua,Alonso:2013hga}.



\bm{$d=8$} The above analysis can be straightforwardly generalized to the renormalization at higher dimensions.
At dimension 8, we have many more types of operators that have multiple Lorentz structures; hence one should diagonalize the $W^2$ operator in the space of basis amplitudes generated by them to get partial wave amplitude basis $\mc{B}^{j,a}$. In case there are degeneracies when we look at a channel with more than two particles, we can further find eigenfunctions for subsets of the channel.
%
In appendix~\ref{sec:W2Diagonalize}, we present the partial wave amplitude basis in the channel $\{1,2,3\}$ and $\{1,2\}$ for the operator types $\psi_1\phi_2\phi_3\psi_4\phi_5D^2$ and $\psi_1 \phi_2\phi_3 \bar\psi_4 F_5 D$, labelled by superscript $(j_{123},j_{12})$.
In this basis, only ($\mathcal{B}_{\psi^2\phi^3D^2}^{3/2,3/2},\mathcal{B}_{F\psi \bar \psi \phi^2D}^{3/2,3/2}$) and ($\mathcal{B}_{\psi^2\phi^3D^2}^{3/2,1/2},\mathcal{B}_{F\psi \bar \psi \phi^2D}^{3/2,1/2}$) can mix with each other through RG running.

Moreover, we find new non-renormalization relations for whole types of operators that were not predicted by~\cite{Cheung:2015aba}.
For example, the $F\bar F \psi \bar \psi D$ operators have $j=2$ in the $(\psi,\bar \psi)$ channel and $\psi \bar\psi\phi^4D$ have $j=1$. Although allowed by the $(\omega,\bar \omega)$ criteria, all $F\bar F \psi \bar \psi D$ type operators can not renormalize the $\psi \bar\psi\phi^4D$ type.

\begin{figure}
\centering
\begin{scaletikzpicturetowidth}{0.5\textwidth}
\begin{tikzpicture}[scale=\tikzscale]

\node[rectangle,fill]  (amp1) at (0,0) {};
\node at (-0.5,0) {$O^{(3)}_{Hl}$};
\node at (2.3,-0.2) {$-4\lambda$};
\node at (1.6,0.2) {$P$};

\draw[thick] (amp1) to [out=45,in=135] (0.6,0) node[label=$$] {};
\draw[thick] (amp1) to [out=-45,in=-135] (0.6,0) node[label=$$] {};
\draw[thick,dashed] (amp1) to [out=-135,in=45] (-0.5,-0.5) node[label=$$] {};
\draw[thick,dashed] (amp1) to [out=135,in=-45] (-0.5,0.5) node[label=$$] {};
\draw[thick,photon] (0.6,0) to [out=0,in=180] (1.2,0) node[label=$$] {};
\draw[thick,dashed] (1.2,0) to [out=90,in=-90] (1.2,0.5) node[label=$$] {};
\draw[thick,dashed] (1.2,0) to [out=0,in=180] (2.5,0) node[label=$$] {};
\draw[thick,dashed] (2,0.5) to [out=-90,in=90] (2,-0.5) node[label=$$] {};

\draw[->] (2.8,0)--(4.2,0);
\node at (3.5,0.2) {\tiny UV divergence};

\begin{scope}[xshift=5.0cm]
\node[rectangle,fill]  (amp1) at (0,0) {};
\node at (-0.5,-0.2) {$H^4D^2$};
\node at (1.3,-0.2) {$-4\lambda$};
\node at (0.5,0.2) {$P$};

\draw[thick,dashed] (-0.5,0) to [out=0,in=180] (1.5,0) node[label=$$] {};
\draw[thick,dashed] (0,0.5) to [out=-90,in=90] (0,-0.5) node[label=$$] {};
\draw[thick,dashed] (1,0.5) to [out=-90,in=90] (1,-0.5) node[label=$$] {};
\end{scope}

\end{tikzpicture}
\end{scaletikzpicturetowidth}
\caption{Renormalization of $O_{H}$ by $O^{(3)}_{Hl}$ at one loop, with $\mc{A}'\neq0$.} \label{fig:RG2}
\end{figure}
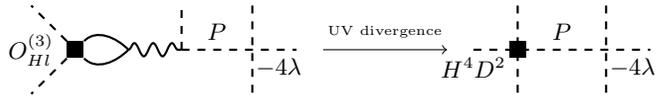

\bi{ $\mc{A}'$ subtlety: } Now we briefly look at an example that has subtlety related to $\mc{A}'$ in  eq.~\eqref{eq:renorm}. Table~\ref{tab:d6RG} shows that the operator $\psi\bar{\psi}^2\phi^2D$ and $\phi^6$ could not renormalize each other. However, in the SMEFT, as shown in \citep{Alonso:2013hga}, $\dot{C}_H \supset \lambda g_2^2 C^3_{Hl}$ has non-vanishing coefficient. The secret hides in the operator basis.
Our selection rule only proves that 1PI diagrams with no $\mc{A}'$ vanishes, thus the non-vanishing diagram for $\mc{A}^{\rm 1-loop}(H^6)$ must be non-1PI as shown in fig.~\ref{fig:RG2}. From angular momentum and isospin of $\mc{O}^3_{Hl}$, we assert that the subamplitude at the LHS of the $P$ propagator should be proportional to the basis amplitude $\mc{B}^{j=1,I=1}_{H^4D^2}$ which renormalizes the operator $(H^\dagger i\tau^i\overleftrightarrow{D}_\mu H)^2$. 
Because it is not Warsaw basis operator, we need to decompose $\mc{B}^{j=1,I=1}_{H^4D^2}$ into those generated by Warsaw basis $O_{H\Box}$ and $O_{HD}$, while we keep in mind that one of the external legs is not on shell $P^2 \neq 0$. 
As a result, we have
\eq{
	\mc{A}^{\rm 1-loop}_{\rm UV}(H^6) &\sim (g_2^2 C^3_{Hl}\mc{B}^{j=1,I=1}_{H^4D^2})\frac{1}{\epsilon}\times\frac{1}{P^2}\times(-4\lambda) \\
	&= \frac{-4\lambda}{P^2}\left( \frac34g_2^2 C^3_{Hl}\mc{B}_{H\Box} - \frac12g_2^2 C^3_{Hl}P^2\right)\frac{1}{\epsilon} \\
	&= \left( 2\lambda g_2^2 C^3_{Hl} - \frac{3\lambda g_2^2 C^3_{Hl}\mc{B}_{H\Box}}{P^2}\right)\frac{1}{\epsilon}.
}
Compared to  eq.~\eqref{eq:renorm}, we see that when $\mc{A}'\neq0$, it is ambiguous to define the ``local'' part of the UV divergence that is supposed to be the anomalous dimension $\gamma$. From angular momentum point of view, $\mc{B}^{j=1,I=1}_{H^4D^2}$ is preferred in the residue of the $P^2$ pole, so that $\gamma=0$ as shown in the second line; but from operator basis point of view, $\mc{B}_{H\Box}$ is the basis contribution to the residue, hence we are left with a local piece for $\dot{C}_H$ proportional to $\lambda g_2^2$ as shown in the third line. 
This computation also gives the correct proportionality
\eq{
\frac{\gamma_{\mc{O}^3_{Hl} \to \mc{O}_{H}}}{\gamma_{\mc{O}^3_{Hl} \to \mc{O}_{H\Box}}} = \frac{2\lambda g_2^2}{\frac34g_2^2} = \frac{8}{3}\lambda
,} 
without any actual loop calculations, 
which agrees with the result in \citep{Alonso:2013hga}.
Given the powerful predictivity, a more systematic way to examine the $\mc{A}'\neq0$ situation needs to be explored.
%

\subsection{Vanishing Loops}

In the previous section, we are considering renormalization, so we have a ``target operator" to be renormalized, which has definite $j$ at the specific channel.
We can also consider the full amplitude, which in general does not have a definite $j$. However, some of them have constrained $j$ at specific channels, which select the operators that contribute to these amplitudes at the loop level.

\begin{figure}
\centering
\begin{scaletikzpicturetowidth}{0.5\textwidth}
\begin{tikzpicture}[scale=\tikzscale]
 \node[rectangle,fill]  (amp1) at (2,0) {};
  \node at (2.8,0) {$F^2\phi^2$};
\node[draw,circle,fill=gray,inner sep=6pt] (amp2) at (0,0) {$d=4$};

\draw[thick,photon] (0.54,0.54) arc (130:10:0.8) ;
\draw[thick,photon] (0.54,-0.54) arc (-130:-10:0.8) ;
\draw[thick,dashed] (amp1) to [out=-45,in=135] (3,-1) node[label=$$] {};
\draw[thick,dashed] (amp1) to [out=45,in=-135] (3,1) node[label=$$] {};
\draw[thick,photon] (amp2) to [out=140,in=-40] (-1.3,1) node[label=$+$] {};
\draw[thick,photon] (amp2) to [out=-140,in=40] (-1.3,-1) node[label=$-$] {};
  \begin{scope}[xshift=6cm]
 \node[rectangle,fill]  (amp1) at (2,0) {};
  \node at (2.8,0) {$F^2\phi^2$};
\node[draw,circle,fill=gray,inner sep=6pt] (amp2) at (0,0) {$d=4$};

\draw[thick,photon] (0.54,0.54) arc (130:10:0.8) ;
\draw[thick,dashed] (0.54,-0.54) arc (-130:-10:0.8) ;
(amp1) to [out=-45,in=180] (1,-0.7) node[label=$$] {};
\draw[thick,photon] (amp1) to [out=-45,in=135] (3,-1) node[label=$+$] {};
\draw[thick,dashed] (amp1) to [out=45,in=-135] (3,1) node[label=$$] {};
\draw[thick,dashed] (amp2) to [out=140,in=-50] (-1.3,1.15) node[label=$$] {};
\draw[thick,photon] (amp2) to [out=-140,in=50] (-1.3,-1) node[label=$-$] {};
\end{scope}
\end{tikzpicture}
\end{scaletikzpicturetowidth}
\caption{Contribution from $F^2\phi^2$ to $\mathcal{A}(F^+F^-\phi \phi)$. The left diagram vanishes because $F^2\phi^2$ excites $j=0$ partial wave for the two external scalars. However, due to the existence of the right diagram, which does not vanish, the full amplitude is non-zero.} \label{fig:nonv}
\end{figure}
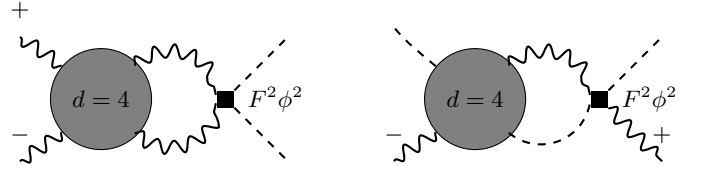

We consider two ways that $j$ can be constrained for two-particle states:
\begin{itemize}
	\item In the Center of Mass (COM) frame we can assure that the orbital angular momentum $\mathbf{r}\times\mathbf{p}$ has vanishing projection along $\hat{p}$, hence along this direction we have $\sigma\equiv J\cdot\hat{p} = S\cdot\hat{p}$, which for massless particles is $\Delta h$, the difference of helicities. Thus we must have $j\geq|\Delta h|$. After covariantizing, $j$ determined by the eigenvalue of $W^2$ satisfies the constraint in any frame.
	This is a generalization of the Weinberg-Witten theorem as noted in \cite{Arkani-Hamed:2017jhn}.


	\item From eq.~\eqref{eq:two_massless}, we find that if $h_1=h_2=h$ the permutation symmetry of the two particles in the CG coefficient is determined by the exponent in $[12]^{j+2h}$. Thus by spin-statistics, once the two particles are identical, the permutation symmetry is allowed only if $j$ is even. It does not apply if there are other group factors that contribute to the permutation symmetry, which is discussed in detail in \citep{ListingOperators:2020}, but for simplicity, we only discuss the example when no group factors exist (also see \citep{Arkani-Hamed:2017jhn}), like for photons, so that all odd $j$s are forbidden.
The Landau-Yang theorem \citep{Landau1948,PhysRev.77.242}, which states that the two-photon state with $j=1$ is forbidden, is nothing but a combination of the above two criteria, the first forbidding the opposite helicity case and the second forbidding the same helicity state.

\end{itemize}
%
Both of the constraints may be used to prove a diagram-level selection rule for the $2\to N$ scattering with an effective operator contributing in the way as shown in fig.~\ref{fig:2to2}.
We should emphasize that the selection rule is only at diagram-level; it is possible for an amplitude to have both vanishing diagrams by the selection rule and non-vanishing ones.
An example is the contribution from $F^2\phi^2$ to $\mathcal{A}(F^+F^-\phi \phi)$ as shown in fig.~\ref{fig:nonv}.


%
\bi{Selection Rule A:} If the two external legs on the LHS of this diagram have helicities differ by $\Delta h $, it then selects the $j\geq \Delta h$ partial waves also for the RHS state, thus forbidding the contribution from an operator with lower $j$ in the specific channel. Such a selection rule may be non-trivial when $\Delta h\geq1$.

At $d=6$, it can be verified that no effective operators can excite two-particle states with $j>1$. So all the diagrams like fig.~\ref{fig:2to2} with $(F^+,F^-)$ or $(F^\pm,\psi^\mp)$ at the LHS and dimension 6 operators inserted at the RHS must vanish. At higher dimensions, non-vanishing contributions will show up, mostly because adding more derivatives to an operator makes it possible to generate higher angular momentum partial waves.

\begin{figure}
\centering
\begin{tikzpicture}
 \node[rectangle,fill]  (amp1) at (2,0) {};
\node[draw,circle,fill=gray,inner sep=6pt] (amp2) at (0,0) {$d=4$};
\fill (2.6,0.3) circle (2pt);
\fill (2.5,0) circle (2pt);
\fill (2.6,-0.3) circle (2pt);
\draw[thick] (amp2) to [out=45,in=180] (1,0.7) node[label=$$] {};
\draw[thick] (amp2) to [out=-45,in=180] (1,-0.7) node[label=$$] {};
\draw[thick] (amp1) to [out=135,in=0] (1,0.7) node[label=$$] {};
\draw[thick] (amp1) to [out=-135,in=0] (1,-0.7) node[label=$$] {};
\draw[thick] (amp2) to [out=-135,in=45] (-1,-1) node[label=$$] {};
\draw[thick] (amp2) to [out=135,in=-45] (-1,1) node[label=$$] {};
\draw[thick] (amp1) to [out=45,in=-135] (3,1) node[label=$$] {};
\draw[thick] (amp1) to [out=-45,in=135] (3,-1) node[label=$$] {};
\end{tikzpicture}
\caption{One loop diagram for $2\rightarrow N$ scattering in EFT.} \label{fig:2to2}
\end{figure}
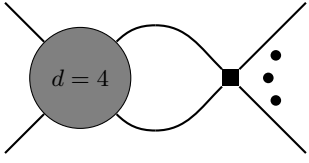

\begin{table}[htbp]
    \centering
    \begin{tabular}{p{1cm}<{\centering}|p{1.2cm}<{\centering}|c|p{3.3cm}<{\centering}}
        \hline
        \hline
        LHS ($\Delta h$) & fields in loop & RHS & EFT operators at RHS \\
        \hline
        $F^+F^-$ (2) & $\phi\phi$ & $\phi\phi$ & $\phi^4$(\textcolor{red}{0}), $\phi^4D^2$(\textcolor{red}{0}, \textcolor{red}{1}), $\phi^4D^4$(\textcolor{red}{0}, \textcolor{red}{1}, 2) \\
        \cline{3-4}
         & & $\psi^+\psi^-$ & $\psi\bar{\psi}\phi^2D^3$(\textcolor{red}{1}), $\psi\bar{\psi}\phi^2D^3$(\textcolor{red}{1}, 2)\\
        \cline{3-4}
         & & $F^+F^+$ & $F^2\phi^2$(\textcolor{red}{0}), $F^2\phi^2D^2$(\textcolor{red}{0}, \textcolor{red}{1})\\
        \cline{2-4}
         & $\psi\bar{\psi}$ & $\phi\phi$ & $\psi\bar{\psi}\phi^2D$(\textcolor{red}{1}), $\psi\bar{\psi}\phi^2D^3$(\textcolor{red}{1}, 2)\\
        \cline{3-4}
         & & $\psi^+\psi^-$ & $\bar{\psi}^2\psi^2$(\textcolor{red}{1}), $\bar{\psi}^2\psi^2D^2$(\textcolor{red}{1}, 2)\\
        \cline{3-4}
         & & $F^+F^+$ & $F^2\psi\bar{\psi}D$(\textcolor{red}{1})\\
        \hline
        $F^+\psi^-$ (3/2) & $\bar{\psi}\phi$ & $\psi^-\phi$ & $\psi\bar{\psi}\phi^2D$(\textcolor{red}{1/2}), $\psi\bar{\psi}\phi^2D^3$(\textcolor{red}{1/2}, 3/2)\\
        \cline{3-4}
         & & $F^+\psi^+$ & $F\psi^2\phi$(\textcolor{red}{1/2}), $F\psi^2\phi D^2$(\textcolor{red}{1/2}, 3/2)\\
        \hline
        $F^+\phi$ (1) & $\psi\psi$ & $\psi^{\pm}\psi^{\pm}$ & $\bar{\psi}^2\psi^2$(\textcolor{red}{0}), $\psi^4$(\textcolor{red}{0},1), $\bar{\psi}^2\psi^2D^2$(\textcolor{red}{0},1), $\psi^4D^2$(\textcolor{red}{0}, 1, 2)\\
        \hline
        $\psi^+\psi^-$ (1) & $\phi\phi$ & $F^{\pm}F^{\pm}$ & $F^2\phi^2$(\textcolor{red}{0}), $F^2\phi^2D^2$(\textcolor{red}{0}, 1)\\
        \cline{2-4}
         & $FF$ & $\phi\phi$ & $F^2\phi^2$(\textcolor{red}{0}), $F^2\phi^2D^2$(\textcolor{red}{0}, 1)\\
        \cline{3-4}
         & & $F^{\pm}F^{\pm}$ & $F^2\bar{F}^2(\textcolor{red}{0}), F^4$(\textcolor{red}{0}, 1, 2)\\
        \hline
        \hline
    \end{tabular}
    \caption{Vanishing one loop amplitudes from contribution of specific operators. In the first column we list the two particle states with a minimum angular momentum $j\ge\Delta h$, and are at LHS of the diagram in fig.~\ref{fig:2to2}. In the third column we list the two particle state excited by effective operators as RHS of diagram in fig.~\ref{fig:2to2}. The combinations of these two column gives the $2\to2$ amplitudes under consideration. In the fourth column are the dimension 6 and dimension 8 operators that have vanishing contributions to these amplitudes, which can be obtained by combining the particles in second and third columns. The numbers in bracket represent the angular momentums of the two-particle states from these operators, and are colored red if they are smaller than $\Delta h$.}
    \label{tab:van}
\end{table}

In Table~\ref{tab:van}, we list all the $2\rightarrow2$ processes for which the one-loop contributions from specific operators to the full amplitude vanish. For dimension 6 cases, this table gives the same results as the ``absent rational terms" found in \cite{Craig:2019wmo}, while the underlying mechanism--angular momentum conservation--is more manifest here. For dimension 8 cases, which are not studied in~\cite{Craig:2019wmo}, we also find two vanishing contributions, namely the contribution from $F^2\phi^2D^2$ to $\mathcal{A}(\phi \phi F^+F^-)$ and $F^2\bar F^2$ to $\mathcal{A}(\psi^+\psi^-F^\pm F^\pm)$.
The other dimension 8 operators in this table can excite partial waves that reach the ``threshold'' of $j=\Delta h$ and thus have non-vanishing contributions to the corresponding processes. For these cases, the amplitude selects specific combinations of the operators that may contribute, which is a strong constraint for EFT phenomenology.
%
For example, the $H^4D^4$ type operators (defined in the appendix) contribute to the amplitude $\mathcal{A}(B^+ B^- H^\alpha H^{\dagger \dot \beta} )$ at one loop level only in the form of the $j=2$ partial wave amplitude basis, with coefficients proportional to the combinations
\eq{
	& C^{2,0}=\frac{1}{6}(C_1^{H^4D^4}+\frac13C_2^{H^4D^4}+C_3^{H^4D^4}), \\	
	& C^{2,1}=\frac{1}{6}(C_1^{H^4D^4}-C_2^{H^4D^4}+C_3^{H^4D^4}), \\
}
while the isospin further selects the first one $C^{2,0}$.

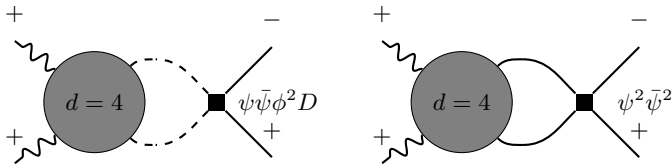
\begin{figure}
\centering
\begin{scaletikzpicturetowidth}{0.5\textwidth}
\begin{tikzpicture}[scale=\tikzscale]
 \node[rectangle,fill]  (amp1) at (2,0) {};
  \node at (3,0) {$\psi\bar{\psi}\phi^2D$};
\node[draw,circle,fill=gray,inner sep=6pt] (amp2) at (0,0) {$d=4$};

\draw[thick,dashed] (amp1) to [out=135,in=0] (1,0.7) node[label=$$] {};
\draw[thick,dashed] (amp1) to [out=-135,in=0] (1,-0.7) node[label=$$] {};
\draw[thick,dashed] (amp2) to [out=45,in=180] (1,0.7) node[label=$$] {};
\draw[thick,dashed] (amp2) to [out=-45,in=180] (1,-0.7) node[label=$$] {};
\draw[thick] (amp1) to [out=-45,in=135] (2.9,-0.9) node[label=$+$] {};
\draw[thick] (amp1) to [out=45,in=-135] (2.9,0.9) node[label=$-$] {};
\draw[thick,photon] (amp2) to [out=140,in=-40] (-1.3,1) node[label=$+$] {};
\draw[thick,photon] (amp2) to [out=-140,in=40] (-1.3,-1) node[label=$+$] {};
  \begin{scope}[xshift=6cm]
 \node[rectangle,fill]  (amp1) at (2,0) {};
  \node at (3,0) {$\psi^2\bar \psi^2$};
\node[draw,circle,fill=gray,inner sep=6pt] (amp2) at (0,0) {$d=4$};

\draw[thick] (amp1) to [out=135,in=0] (1,0.7) node[label=$$] {};
\draw[thick] (amp1) to [out=-135,in=0] (1,-0.7) node[label=$$] {};
\draw[thick] (amp2) to [out=45,in=180] (1,0.7) node[label=$$] {};
\draw[thick] (amp2) to [out=-45,in=180] (1,-0.7) node[label=$$] {};
\draw[thick] (amp1) to [out=-45,in=135] (2.9,-0.9) node[label=$+$] {};
\draw[thick] (amp1) to [out=45,in=-135] (2.9,0.9) node[label=$-$] {};
\draw[thick,photon] (amp2) to [out=140,in=-40] (-1.3,1) node[label=$+$] {};
\draw[thick,photon] (amp2) to [out=-140,in=40] (-1.3,-1) node[label=$+$] {};
  \end{scope}
\end{tikzpicture}
\end{scaletikzpicturetowidth}
\caption{Vanishing loops from $j\ne 1$ for two identical particles of same helicity.} \label{fig:LY}
\end{figure}

\bi{Selection Rule B:} If the LHS state in fig.~\ref{fig:2to2} consists of two identical particles (same gauge charges and same helicities), the $j=1$ partial wave is forbidden on the RHS, selecting the operator that contributes in this specific way.
As an example, consider the one-loop amplitude $\mathcal{A}(\psi^+\psi^-F^+F^+)$ from $\psi \bar \psi \phi^2D$ and $\psi ^2\bar \psi^2$, as in fig.~\ref{fig:LY}, with two gauge bosons being identical, like $B$ in the SM. Because the other two-fermion state created by these two effective operators have exactly $j=1$, this amplitude must vanish.

\section{Conclusion}
In this letter, we have derived the CG coefficients of Poincar\'e group in terms of spinor helicity variables. We showed that they are the normalized amplitude basis between the multi-particle state and an auxiliary massive particle state, the spin of which gives the total angular momentum. Then we obtain the partial wave amplitude basis $\mc{B}^j$ with definite angular momentum, where the $2\to 2$ partial wave amplitude basis reproduces the famous Wigner d-matrix. This allows us to constrain the UV physics that generates an effective operator via tree diagrams. We further develop the technique to get the angular momentum of a generic basis amplitude systematically via the Casimir invariant operator $W^2$ in the spinor helicity representation. 

More importantly, these techniques allow us to find new selection rules based on angular momentum conservation. By using the operator-amplitude basis correspondence, we assign an angular momentum to the operator corresponding to a partial wave amplitude basis in a certain channel. When inserted into one-loop diagrams, either in calculating the renormalization of effective operators or in the loop calculation of full amplitudes, only the operators with correct angular momentum would be selected. 
First, we show how such selection rules are applied to predict new zeros and non-trivial proportionalities in the anomalous dimension matrix of effective operators. Second, we prove two kinds of constraints on the total angular momentum for two-massless-particle states as the analog of the famous Weinberg-Witten and Landau-Yang theorems, both of which prevent the operators with wrong angular momentum from contributing to the $2\to N$ amplitudes by the conservation law.

\section{Acknowledgements}
J.S. is supported by the National Natural Science Foundation of China (NSFC) under grant No.11947302, No.11690022, No.11851302, No.11675243 and No.11761141011 and also supported by the Strategic Priority Research Program of the Chinese Academy of Sciences under grant No.XDB21010200 and No.XDB23000000. M.L.X. is supported by the National Natural Science Foundation of China (NSFC) under grant No.2019M650856 and the 2019 International Postdoctoral Exchange Fellowship Program.

\bibliographystyle{apsrev4-1}
\bibliography{ref}

\begin{thebibliography}{28}%
\makeatletter
\providecommand \@ifxundefined [1]{%
 \@ifx{#1\undefined}
}%
\providecommand \@ifnum [1]{%
 \ifnum #1\expandafter \@firstoftwo
 \else \expandafter \@secondoftwo
 \fi
}%
\providecommand \@ifx [1]{%
 \ifx #1\expandafter \@firstoftwo
 \else \expandafter \@secondoftwo
 \fi
}%
\providecommand \natexlab [1]{#1}%
\providecommand \enquote  [1]{``#1''}%
\providecommand \bibnamefont  [1]{#1}%
\providecommand \bibfnamefont [1]{#1}%
\providecommand \citenamefont [1]{#1}%
\providecommand \href@noop [0]{\@secondoftwo}%
\providecommand \href [0]{\begingroup \@sanitize@url \@href}%
\providecommand \@href[1]{\@@startlink{#1}\@@href}%
\providecommand \@@href[1]{\endgroup#1\@@endlink}%
\providecommand \@sanitize@url [0]{\catcode `\\12\catcode `\$12\catcode
  `\&12\catcode `\#12\catcode `\^12\catcode `\_12\catcode `\%12\relax}%
\providecommand \@@startlink[1]{}%
\providecommand \@@endlink[0]{}%
\providecommand \url  [0]{\begingroup\@sanitize@url \@url }%
\providecommand \@url [1]{\endgroup\@href {#1}{\urlprefix }}%
\providecommand \urlprefix  [0]{URL }%
\providecommand \Eprint [0]{\href }%
\providecommand \doibase [0]{http://dx.doi.org/}%
\providecommand \selectlanguage [0]{\@gobble}%
\providecommand \bibinfo  [0]{\@secondoftwo}%
\providecommand \bibfield  [0]{\@secondoftwo}%
\providecommand \translation [1]{[#1]}%
\providecommand \BibitemOpen [0]{}%
\providecommand \bibitemStop [0]{}%
\providecommand \bibitemNoStop [0]{.\EOS\space}%
\providecommand \EOS [0]{\spacefactor3000\relax}%
\providecommand \BibitemShut  [1]{\csname bibitem#1\endcsname}%
\let\auto@bib@innerbib\@empty
\bibitem [{\citenamefont {Noether}(1918)}]{Noether:1918zz}%
  \BibitemOpen
  \bibfield  {author} {\bibinfo {author} {\bibfnamefont {E.}~\bibnamefont
  {Noether}},\ }\href {\doibase 10.1080/00411457108231446} {\bibfield
  {journal} {\bibinfo  {journal} {Gott. Nachr.}\ }\textbf {\bibinfo {volume}
  {1918}},\ \bibinfo {pages} {235} (\bibinfo {year} {1918})},\ \bibinfo {note}
  {[Transp. Theory Statist. Phys.1,186(1971)]},\ \Eprint
  {http://arxiv.org/abs/physics/0503066} {arXiv:physics/0503066 [physics]}
  \BibitemShut {NoStop}%
\bibitem [{\citenamefont {Yang}(1950)}]{PhysRev.77.242}%
  \BibitemOpen
  \bibfield  {author} {\bibinfo {author} {\bibfnamefont {C.~N.}\ \bibnamefont
  {Yang}},\ }\href {\doibase 10.1103/PhysRev.77.242} {\bibfield  {journal}
  {\bibinfo  {journal} {Phys. Rev.}\ }\textbf {\bibinfo {volume} {77}},\
  \bibinfo {pages} {242} (\bibinfo {year} {1950})}\BibitemShut {NoStop}%
\bibitem [{\citenamefont {Landau}(1948)}]{Landau1948}%
  \BibitemOpen
  \bibfield  {author} {\bibinfo {author} {\bibfnamefont {L.~D.}\ \bibnamefont
  {Landau}},\ }\href@noop {} {\bibfield  {journal} {\bibinfo  {journal} {Dokl.
  Akad. Nauk SSSR.}\ }\textbf {\bibinfo {volume} {60}},\ \bibinfo {pages} {207}
  (\bibinfo {year} {1948})}\BibitemShut {NoStop}%
\bibitem [{\citenamefont {Shadmi}\ and\ \citenamefont
  {Weiss}(2019)}]{Shadmi:2018xan}%
  \BibitemOpen
  \bibfield  {author} {\bibinfo {author} {\bibfnamefont {Y.}~\bibnamefont
  {Shadmi}}\ and\ \bibinfo {author} {\bibfnamefont {Y.}~\bibnamefont {Weiss}},\
  }\href {\doibase 10.1007/JHEP02(2019)165} {\bibfield  {journal} {\bibinfo
  {journal} {JHEP}\ }\textbf {\bibinfo {volume} {02}},\ \bibinfo {pages} {165}
  (\bibinfo {year} {2019})},\ \Eprint {http://arxiv.org/abs/1809.09644}
  {arXiv:1809.09644 [hep-ph]} \BibitemShut {NoStop}%
\bibitem [{\citenamefont {Ma}\ \emph {et~al.}(2019)\citenamefont {Ma},
  \citenamefont {Shu},\ and\ \citenamefont {Xiao}}]{Ma:2019gtx}%
  \BibitemOpen
  \bibfield  {author} {\bibinfo {author} {\bibfnamefont {T.}~\bibnamefont
  {Ma}}, \bibinfo {author} {\bibfnamefont {J.}~\bibnamefont {Shu}}, \ and\
  \bibinfo {author} {\bibfnamefont {M.-L.}\ \bibnamefont {Xiao}},\ }\href@noop
  {} {\  (\bibinfo {year} {2019})},\ \Eprint {http://arxiv.org/abs/1902.06752}
  {arXiv:1902.06752 [hep-ph]} \BibitemShut {NoStop}%
\bibitem [{\citenamefont {Henning}\ and\ \citenamefont
  {Melia}(2019)}]{Henning:2019enq}%
  \BibitemOpen
  \bibfield  {author} {\bibinfo {author} {\bibfnamefont {B.}~\bibnamefont
  {Henning}}\ and\ \bibinfo {author} {\bibfnamefont {T.}~\bibnamefont
  {Melia}},\ }\href {\doibase 10.1103/PhysRevD.100.016015} {\bibfield
  {journal} {\bibinfo  {journal} {Phys. Rev.}\ }\textbf {\bibinfo {volume}
  {D100}},\ \bibinfo {pages} {016015} (\bibinfo {year} {2019})},\ \Eprint
  {http://arxiv.org/abs/1902.06754} {arXiv:1902.06754 [hep-ph]} \BibitemShut
  {NoStop}%
\bibitem [{\citenamefont {Durieux}\ and\ \citenamefont
  {Machado}(2019)}]{Durieux:2019siw}%
  \BibitemOpen
  \bibfield  {author} {\bibinfo {author} {\bibfnamefont {G.}~\bibnamefont
  {Durieux}}\ and\ \bibinfo {author} {\bibfnamefont {C.~S.}\ \bibnamefont
  {Machado}},\ }\href@noop {} {\  (\bibinfo {year} {2019})},\ \Eprint
  {http://arxiv.org/abs/1912.08827} {arXiv:1912.08827 [hep-ph]} \BibitemShut
  {NoStop}%
\bibitem [{\citenamefont {Falkowski}(2019)}]{Falkowski:2019zdo}%
  \BibitemOpen
  \bibfield  {author} {\bibinfo {author} {\bibfnamefont {A.}~\bibnamefont
  {Falkowski}},\ }\href@noop {} {\  (\bibinfo {year} {2019})},\ \Eprint
  {http://arxiv.org/abs/1912.07865} {arXiv:1912.07865 [hep-ph]} \BibitemShut
  {NoStop}%
\bibitem [{\citenamefont {Elias-Miro}\ \emph {et~al.}(2015)\citenamefont
  {Elias-Miro}, \citenamefont {Espinosa},\ and\ \citenamefont
  {Pomarol}}]{Elias-Miro:2014eia}%
  \BibitemOpen
  \bibfield  {author} {\bibinfo {author} {\bibfnamefont {J.}~\bibnamefont
  {Elias-Miro}}, \bibinfo {author} {\bibfnamefont {J.~R.}\ \bibnamefont
  {Espinosa}}, \ and\ \bibinfo {author} {\bibfnamefont {A.}~\bibnamefont
  {Pomarol}},\ }\href {\doibase 10.1016/j.physletb.2015.05.056} {\bibfield
  {journal} {\bibinfo  {journal} {Phys. Lett.}\ }\textbf {\bibinfo {volume}
  {B747}},\ \bibinfo {pages} {272} (\bibinfo {year} {2015})},\ \Eprint
  {http://arxiv.org/abs/1412.7151} {arXiv:1412.7151 [hep-ph]} \BibitemShut
  {NoStop}%
\bibitem [{\citenamefont {Jenkins}\ \emph {et~al.}(2013)\citenamefont
  {Jenkins}, \citenamefont {Manohar},\ and\ \citenamefont
  {Trott}}]{Jenkins:2013zja}%
  \BibitemOpen
  \bibfield  {author} {\bibinfo {author} {\bibfnamefont {E.~E.}\ \bibnamefont
  {Jenkins}}, \bibinfo {author} {\bibfnamefont {A.~V.}\ \bibnamefont
  {Manohar}}, \ and\ \bibinfo {author} {\bibfnamefont {M.}~\bibnamefont
  {Trott}},\ }\href {\doibase 10.1007/JHEP10(2013)087} {\bibfield  {journal}
  {\bibinfo  {journal} {JHEP}\ }\textbf {\bibinfo {volume} {10}},\ \bibinfo
  {pages} {087} (\bibinfo {year} {2013})},\ \Eprint
  {http://arxiv.org/abs/1308.2627} {arXiv:1308.2627 [hep-ph]} \BibitemShut
  {NoStop}%
\bibitem [{\citenamefont {Alonso}\ \emph
  {et~al.}(2014{\natexlab{a}})\citenamefont {Alonso}, \citenamefont {Jenkins},\
  and\ \citenamefont {Manohar}}]{Alonso:2014rga}%
  \BibitemOpen
  \bibfield  {author} {\bibinfo {author} {\bibfnamefont {R.}~\bibnamefont
  {Alonso}}, \bibinfo {author} {\bibfnamefont {E.~E.}\ \bibnamefont {Jenkins}},
  \ and\ \bibinfo {author} {\bibfnamefont {A.~V.}\ \bibnamefont {Manohar}},\
  }\href {\doibase 10.1016/j.physletb.2014.10.045} {\bibfield  {journal}
  {\bibinfo  {journal} {Phys. Lett.}\ }\textbf {\bibinfo {volume} {B739}},\
  \bibinfo {pages} {95} (\bibinfo {year} {2014}{\natexlab{a}})},\ \Eprint
  {http://arxiv.org/abs/1409.0868} {arXiv:1409.0868 [hep-ph]} \BibitemShut
  {NoStop}%
\bibitem [{\citenamefont {Bern}\ \emph {et~al.}(2019)\citenamefont {Bern},
  \citenamefont {Sawyer},\ and\ \citenamefont {Parra-Martinez}}]{Bern:2019wie}%
  \BibitemOpen
  \bibfield  {author} {\bibinfo {author} {\bibfnamefont {Z.}~\bibnamefont
  {Bern}}, \bibinfo {author} {\bibfnamefont {E.}~\bibnamefont {Sawyer}}, \ and\
  \bibinfo {author} {\bibfnamefont {J.}~\bibnamefont {Parra-Martinez}},\
  }\href@noop {} {\  (\bibinfo {year} {2019})},\ \Eprint
  {http://arxiv.org/abs/1910.05831} {arXiv:1910.05831 [hep-ph]} \BibitemShut
  {NoStop}%
\bibitem [{\citenamefont {Cheung}\ and\ \citenamefont
  {Shen}(2015)}]{Cheung:2015aba}%
  \BibitemOpen
  \bibfield  {author} {\bibinfo {author} {\bibfnamefont {C.}~\bibnamefont
  {Cheung}}\ and\ \bibinfo {author} {\bibfnamefont {C.-H.}\ \bibnamefont
  {Shen}},\ }\href {\doibase 10.1103/PhysRevLett.115.071601} {\bibfield
  {journal} {\bibinfo  {journal} {Phys. Rev. Lett.}\ }\textbf {\bibinfo
  {volume} {115}},\ \bibinfo {pages} {071601} (\bibinfo {year} {2015})},\
  \Eprint {http://arxiv.org/abs/1505.01844} {arXiv:1505.01844 [hep-ph]}
  \BibitemShut {NoStop}%
\bibitem [{\citenamefont {Azatov}\ \emph {et~al.}(2017)\citenamefont {Azatov},
  \citenamefont {Contino}, \citenamefont {Machado},\ and\ \citenamefont
  {Riva}}]{Azatov:2016sqh}%
  \BibitemOpen
  \bibfield  {author} {\bibinfo {author} {\bibfnamefont {A.}~\bibnamefont
  {Azatov}}, \bibinfo {author} {\bibfnamefont {R.}~\bibnamefont {Contino}},
  \bibinfo {author} {\bibfnamefont {C.~S.}\ \bibnamefont {Machado}}, \ and\
  \bibinfo {author} {\bibfnamefont {F.}~\bibnamefont {Riva}},\ }\href {\doibase
  10.1103/PhysRevD.95.065014} {\bibfield  {journal} {\bibinfo  {journal} {Phys.
  Rev.}\ }\textbf {\bibinfo {volume} {D95}},\ \bibinfo {pages} {065014}
  (\bibinfo {year} {2017})},\ \Eprint {http://arxiv.org/abs/1607.05236}
  {arXiv:1607.05236 [hep-ph]} \BibitemShut {NoStop}%
\bibitem [{\citenamefont {Craig}\ \emph {et~al.}(2019)\citenamefont {Craig},
  \citenamefont {Jiang}, \citenamefont {Li},\ and\ \citenamefont
  {Sutherland}}]{Craig:2019wmo}%
  \BibitemOpen
  \bibfield  {author} {\bibinfo {author} {\bibfnamefont {N.}~\bibnamefont
  {Craig}}, \bibinfo {author} {\bibfnamefont {M.}~\bibnamefont {Jiang}},
  \bibinfo {author} {\bibfnamefont {Y.-Y.}\ \bibnamefont {Li}}, \ and\ \bibinfo
  {author} {\bibfnamefont {D.}~\bibnamefont {Sutherland}},\ }\href@noop {} {\
  (\bibinfo {year} {2019})},\ \Eprint {http://arxiv.org/abs/2001.00017}
  {arXiv:2001.00017 [hep-ph]} \BibitemShut {NoStop}%
\bibitem [{Note1()}]{Note1}%
  \BibitemOpen
  \bibinfo {note} {For more complicated cases, one needs to use the reduced
  Young Tableau~\cite {Henning:2019enq} or momentum twistors~\cite
  {Falkowski:2019zdo} to deal with momentum conservation.}\BibitemShut {Stop}%
\bibitem [{\citenamefont {Grzadkowski}\ \emph {et~al.}(2010)\citenamefont
  {Grzadkowski}, \citenamefont {Iskrzynski}, \citenamefont {Misiak},\ and\
  \citenamefont {Rosiek}}]{Grzadkowski:2010es}%
  \BibitemOpen
  \bibfield  {author} {\bibinfo {author} {\bibfnamefont {B.}~\bibnamefont
  {Grzadkowski}}, \bibinfo {author} {\bibfnamefont {M.}~\bibnamefont
  {Iskrzynski}}, \bibinfo {author} {\bibfnamefont {M.}~\bibnamefont {Misiak}},
  \ and\ \bibinfo {author} {\bibfnamefont {J.}~\bibnamefont {Rosiek}},\ }\href
  {\doibase 10.1007/JHEP10(2010)085} {\bibfield  {journal} {\bibinfo  {journal}
  {JHEP}\ }\textbf {\bibinfo {volume} {10}},\ \bibinfo {pages} {085} (\bibinfo
  {year} {2010})},\ \Eprint {http://arxiv.org/abs/1008.4884} {arXiv:1008.4884
  [hep-ph]} \BibitemShut {NoStop}%
\bibitem [{\citenamefont {Arkani-Hamed}\ \emph {et~al.}(2017)\citenamefont
  {Arkani-Hamed}, \citenamefont {Huang},\ and\ \citenamefont
  {Huang}}]{Arkani-Hamed:2017jhn}%
  \BibitemOpen
  \bibfield  {author} {\bibinfo {author} {\bibfnamefont {N.}~\bibnamefont
  {Arkani-Hamed}}, \bibinfo {author} {\bibfnamefont {T.-C.}\ \bibnamefont
  {Huang}}, \ and\ \bibinfo {author} {\bibfnamefont {Y.-t.}\ \bibnamefont
  {Huang}},\ }\href@noop {} {\  (\bibinfo {year} {2017})},\ \Eprint
  {http://arxiv.org/abs/1709.04891} {arXiv:1709.04891 [hep-th]} \BibitemShut
  {NoStop}%
\bibitem [{Note2()}]{Note2}%
  \BibitemOpen
  \bibinfo {note} {Here ``unfactorizable'' means that the amplitude does not
  have any kinematic poles or branch cuts on which it could be factorized by
  unitarity.}\BibitemShut {Stop}%
\bibitem [{Note3()}]{Note3}%
  \BibitemOpen
  \bibinfo {note} {The partial wave expansion for long-range scattering is
  tricky, which involves zero-poles in other channels and an infinite tower of
  $j$ in the summation. In this letter, we temporarily ignore such amplitudes,
  and mainly focus on the basis amplitudes.}\BibitemShut {Stop}%
\bibitem [{\citenamefont {Arzt}\ \emph {et~al.}(1995)\citenamefont {Arzt},
  \citenamefont {Einhorn},\ and\ \citenamefont {Wudka}}]{Arzt:1994gp}%
  \BibitemOpen
  \bibfield  {author} {\bibinfo {author} {\bibfnamefont {C.}~\bibnamefont
  {Arzt}}, \bibinfo {author} {\bibfnamefont {M.~B.}\ \bibnamefont {Einhorn}}, \
  and\ \bibinfo {author} {\bibfnamefont {J.}~\bibnamefont {Wudka}},\ }\href
  {\doibase 10.1016/0550-3213(94)00336-D} {\bibfield  {journal} {\bibinfo
  {journal} {Nucl. Phys.}\ }\textbf {\bibinfo {volume} {B433}},\ \bibinfo
  {pages} {41} (\bibinfo {year} {1995})},\ \Eprint
  {http://arxiv.org/abs/hep-ph/9405214} {arXiv:hep-ph/9405214 [hep-ph]}
  \BibitemShut {NoStop}%
\bibitem [{\citenamefont {de~Blas}\ \emph {et~al.}(2018)\citenamefont
  {de~Blas}, \citenamefont {Criado}, \citenamefont {Perez-Victoria},\ and\
  \citenamefont {Santiago}}]{deBlas:2017xtg}%
  \BibitemOpen
  \bibfield  {author} {\bibinfo {author} {\bibfnamefont {J.}~\bibnamefont
  {de~Blas}}, \bibinfo {author} {\bibfnamefont {J.~C.}\ \bibnamefont {Criado}},
  \bibinfo {author} {\bibfnamefont {M.}~\bibnamefont {Perez-Victoria}}, \ and\
  \bibinfo {author} {\bibfnamefont {J.}~\bibnamefont {Santiago}},\ }\href
  {\doibase 10.1007/JHEP03(2018)109} {\bibfield  {journal} {\bibinfo  {journal}
  {JHEP}\ }\textbf {\bibinfo {volume} {03}},\ \bibinfo {pages} {109} (\bibinfo
  {year} {2018})},\ \Eprint {http://arxiv.org/abs/1711.10391} {arXiv:1711.10391
  [hep-ph]} \BibitemShut {NoStop}%
\bibitem [{\citenamefont {Witten}(2004)}]{Witten:2003nn}%
  \BibitemOpen
  \bibfield  {author} {\bibinfo {author} {\bibfnamefont {E.}~\bibnamefont
  {Witten}},\ }\href {\doibase 10.1007/s00220-004-1187-3} {\bibfield  {journal}
  {\bibinfo  {journal} {Commun. Math. Phys.}\ }\textbf {\bibinfo {volume}
  {252}},\ \bibinfo {pages} {189} (\bibinfo {year} {2004})},\ \Eprint
  {http://arxiv.org/abs/hep-th/0312171} {arXiv:hep-th/0312171 [hep-th]}
  \BibitemShut {NoStop}%
\bibitem [{Note4()}]{Note4}%
  \BibitemOpen
  \bibinfo {note} {We are using a slightly different normalization than that
  in~\protect \citep {Witten:2003nn}.}\BibitemShut {Stop}%
\bibitem [{Note5()}]{Note5}%
  \BibitemOpen
  \bibinfo {note} {We use $(\phi , \psi _\alpha , \protect \mathaccentV
  {bar}016{\psi }_{\protect \mathaccentV {dot}05F\alpha }, F_{\alpha \beta
  },\protect \mathaccentV {bar}016F_{\protect \mathaccentV {dot}05F\alpha
  \protect \mathaccentV {dot}05F\beta })$ to denote generic fields transforming
  under Lorentz group $SU(2)_L \times SU(2)_R \equiv SO(3,1)$ as (0,0),
  (1/2,0), (0,1/2), (1,0) and (0,1). And $D$ is the covariant derivative. We
  also use $(\phi ,\psi ^\pm ,F^\pm )$ to denote on-shell scalar, fermion, and
  vector particles with $\pm $ helicities in scattering
  amplitudes.}\BibitemShut {Stop}%
\bibitem [{\citenamefont {Jenkins}\ \emph {et~al.}(2014)\citenamefont
  {Jenkins}, \citenamefont {Manohar},\ and\ \citenamefont
  {Trott}}]{Jenkins:2013wua}%
  \BibitemOpen
  \bibfield  {author} {\bibinfo {author} {\bibfnamefont {E.~E.}\ \bibnamefont
  {Jenkins}}, \bibinfo {author} {\bibfnamefont {A.~V.}\ \bibnamefont
  {Manohar}}, \ and\ \bibinfo {author} {\bibfnamefont {M.}~\bibnamefont
  {Trott}},\ }\href {\doibase 10.1007/JHEP01(2014)035} {\bibfield  {journal}
  {\bibinfo  {journal} {JHEP}\ }\textbf {\bibinfo {volume} {01}},\ \bibinfo
  {pages} {035} (\bibinfo {year} {2014})},\ \Eprint
  {http://arxiv.org/abs/1310.4838} {arXiv:1310.4838 [hep-ph]} \BibitemShut
  {NoStop}%
\bibitem [{\citenamefont {Alonso}\ \emph
  {et~al.}(2014{\natexlab{b}})\citenamefont {Alonso}, \citenamefont {Jenkins},
  \citenamefont {Manohar},\ and\ \citenamefont {Trott}}]{Alonso:2013hga}%
  \BibitemOpen
  \bibfield  {author} {\bibinfo {author} {\bibfnamefont {R.}~\bibnamefont
  {Alonso}}, \bibinfo {author} {\bibfnamefont {E.~E.}\ \bibnamefont {Jenkins}},
  \bibinfo {author} {\bibfnamefont {A.~V.}\ \bibnamefont {Manohar}}, \ and\
  \bibinfo {author} {\bibfnamefont {M.}~\bibnamefont {Trott}},\ }\href
  {\doibase 10.1007/JHEP04(2014)159} {\bibfield  {journal} {\bibinfo  {journal}
  {JHEP}\ }\textbf {\bibinfo {volume} {04}},\ \bibinfo {pages} {159} (\bibinfo
  {year} {2014}{\natexlab{b}})},\ \Eprint {http://arxiv.org/abs/1312.2014}
  {arXiv:1312.2014 [hep-ph]} \BibitemShut {NoStop}%
\bibitem [{\citenamefont {Li}\ \emph {et~al.}()\citenamefont {Li},
  \citenamefont {Ren}, \citenamefont {Shu}, \citenamefont {Xiao}, \citenamefont
  {Yu},\ and\ \citenamefont {Zheng}}]{ListingOperators:2020}%
  \BibitemOpen
  \bibfield  {author} {\bibinfo {author} {\bibfnamefont {H.}~\bibnamefont
  {Li}}, \bibinfo {author} {\bibfnamefont {Z.}~\bibnamefont {Ren}}, \bibinfo
  {author} {\bibfnamefont {J.}~\bibnamefont {Shu}}, \bibinfo {author}
  {\bibfnamefont {M.-L.}\ \bibnamefont {Xiao}}, \bibinfo {author}
  {\bibfnamefont {J.-H.}\ \bibnamefont {Yu}}, \ and\ \bibinfo {author}
  {\bibfnamefont {Y.-H.}\ \bibnamefont {Zheng}},\ }\href@noop {} {\bibinfo
  {journal} {work in progress}\ }\BibitemShut {NoStop}%
\end{thebibliography}%

\appendix


\section{Relevant effective operators in SMEFT}
\label{sec:d6ops}
We list the the dimension 6 operators in Warsaw basis \citep{Grzadkowski:2010es} relevant in the main text in table~\ref{operatorbasis}.
\begin{table}[htbp]
    \centering
    \begin{tabular}{|c|c||c|c|}
        \hline
        \multicolumn{2}{|c||}{$\psi^2H^3$} & \multicolumn{2}{c|}{$\psi^4$}\\\hline
        $\mathcal{O}_{eH}$ & $H^\dagger H(\bar{l}eH)$ &	$\mathcal{O}^1_{lequ}$ & $(\bar{l}e)\epsilon_{jk}(\bar{q}u)$ \\\cline{1-2}
		\multicolumn{2}{|c||}{$\psi^2H^2D$} & $\mathcal{O}^3_{lequ}$ & $(\bar{l}\sigma_{\mu\nu}e)\epsilon_{jk}(\bar{q}\sigma^{\mu\nu}u)$ \\\cline{1-4}
        $\mathcal{O}^1_{Hl}$ & $(H^\dagger i\overleftrightarrow{D}_{\mu}H)(\bar{l}\gamma^{\mu}l)$ & \multicolumn{2}{c|}{$\psi^2XH$} \\\cline{3-4}
        $\mathcal{O}^3_{Hl}$ & $(H^\dagger i\overleftrightarrow{D}_{\mu}\tau^aH)(\bar{l}\gamma^{\mu}\tau^al)$ & $\mathcal{O}_{eW}$ & $(\bar{l}\sigma^{\mu\nu}e)\tau^aHW^a_{\mu\nu}$ \\
        $\mathcal{O}_{He}$ & $(H^\dagger i\overleftrightarrow{D}_{\mu}H)(\bar{e}\gamma^{\mu}e)$ & $\mathcal{O}_{eB}$ & $(\bar{l}\sigma^{\mu\nu}e)HB_{\mu\nu}$ \\\cline{3-4}
        $\mathcal{O}^1_{Hq}$ & $(H^\dagger i\overleftrightarrow{D}_{\mu}H)(\bar{q}\gamma^{\mu}q)$ & \multicolumn{2}{c|}{$H^4D^2$} \\\cline{3-4}
        $\mathcal{O}^3_{Hq}$ & $(H^\dagger i\overleftrightarrow{D}_{\mu}\tau^aH)(\bar{q}\gamma^{\mu}\tau^aq)$ & $\mathcal{O}_{H\square}$ & $(H^\dagger H)\square(H^\dagger H)$\\
        $\mathcal{O}_{Hu}$ & $(H^\dagger i\overleftrightarrow{D}_{\mu}H)(\bar{u}\gamma^{\mu}u)$ & $\mathcal{O}_{HD}$ & $(H^\dagger D_{\mu}H)^*(H^\dagger D^{\mu}H)$ \\\cline{3-4}
        $\mathcal{O}_{Hd}$ & $(H^\dagger i\overleftrightarrow{D}_{\mu}H)(\bar{d}\gamma^{\mu}d)$ & \multicolumn{2}{c|}{$H^6$} \\\cline{3-4}
		& & $\mathcal{O}_{H}$ & $(H^\dagger H)^3$ \\
    	\hline
    \end{tabular}
    \caption{Relevant dimension 6 operators in SMEFT.}\label{operatorbasis}
\end{table}

For the dimension 8 $\phi^4D^4$ type operators, we write the corresponding amplitude basis as
\bea
\mathcal{B}_{1}(H^{\alpha} H^{\beta} H^{\dagger \dot \alpha} H^{\dagger \dot \beta}) &=&   (\delta^{\alpha \dot\alpha} \delta^{\beta \dot\beta} + \delta^{\beta \dot\alpha} \delta^{\alpha \dot\beta}) (s_{13} -s_{23})^{2},\nonumber\\
\mathcal{B}_{2}(H^{\alpha} H^{\beta} H^{\dagger \dot \alpha} H^{\dagger \dot \beta}) &=& (\delta^{\alpha \dot\alpha} \delta^{\beta \dot\beta}  - \delta^{\beta \dot\alpha} \delta^{\alpha \dot\beta}) s_{12} (s_{13} -s_{23}),\nonumber \\
\mathcal{B}_{3}(H^{\alpha} H^{\beta} H^{\dagger \dot \alpha} H^{\dagger \dot \beta})&=&(\delta^{\alpha \dot\alpha} \delta^{\beta \dot\beta}  + \delta^{\beta \dot\alpha} \delta^{\alpha \dot\beta})s_{12}^2.
\label{eq:B_H^4D^4}
\eea
while the Wilson coefficients are denoted as $C_i^{H^4D^4}$.

\section{Diagonalization of $W^2$ matrix}
\label{sec:W2Diagonalize}

Since $W^2$ commutes with all Poincar\'e generators and dilatation, it has a matrix representation in the space of basis amplitudes with the same particle states and dimension, whose basis can be described by the reduced Semi-Standard Young Tableau (rSSYT) \citep{Henning:2019enq}. Therefore we can simply diagonalize it and obtain a partial wave amplitude basis. We have realized the $W^2$ operator in \textsf{Mathematica}, which quickly does the diagonalization.

To present a non-trivial result, we take the operator $\psi^2\phi^3D^2$ as an example, which has 6 terms regardless of group factor. We choose the channel $\psi_1\phi_2\phi_3 \to \psi_4\phi_5$, where we first classify the basis amplitudes in terms of $j_{123}$. The algorithm goes as follows
\begin{itemize}
	\item Find an initial amplitude basis $\mc{B}_i, i=1,\dots,6$ using the rSSYT. In this case we get
	\eq{
	\begin{array}{|c|c|c|}
		\hline\hline
		i	&	\mc{B}_i				&	\mc{O}_i	\\
		\hline
		1	&	\vev{12}\vev{45}[25]	&	-\psi_1\sigma^\mu\bar\sigma^\nu\psi_4(D_\mu\phi_2)\phi_3(D_\nu\phi_5)\\
		2	&	\vev{12}\vev{34}[23]	&	\psi_1\sigma^\mu\bar\sigma^\nu\psi_4(D_\mu\phi_2)(D_\nu\phi_3)\phi_5\\
		3	&	\vev{15}\vev{34}[35]	&	-\psi_1\sigma^\mu\bar\sigma^\nu\psi_4\phi_2(D_\nu\phi_3)(D_\mu\phi_5)\\
		4	&	\vev{14}\vev{25}[25]	&	-\psi_1\psi_4(D_\mu\phi_2)\psi_3(D^\mu\phi_5)\\
		5	&	\vev{14}\vev{23}[23]	&	-\psi_1\psi_4(D_\mu\phi_2)(D^\mu\phi_3)\phi_5\\
		6	&	\vev{14}\vev{35}[35]	&	-\psi_1\psi_4\phi_2(D_\mu\phi_3)(D^\mu\phi_5)\\
		\hline\hline
	\end{array}
	}
	
	\item Applying $W^2$ to them according to eq.~\eqref{eq:Casimir_W2}, and get the coefficient matrix $W^2\mc{B}_i = \sum_j\mc{W}_{ij}s_{123}\mc{B}_j$
	\eq{
		\mc{W} = \begin{pmatrix}
			-\dfrac34	&0	&0	&2	&0	&0 \\
			0	&-\dfrac34	&1	&-1	&0	&1 \\
			0	&0	&-\dfrac{11}4&0	&0	&-2	\\
			0	&0	&0	&-\dfrac{15}4&0	&0 \\
			0	&0	&0	&-1	&-\dfrac34	&0 \\
			0	&0	&-1	&0	&0	&-\dfrac74
		\end{pmatrix}.
	}
	
	\item Diagonalizing $\mc{W}$ to obtain the eigenvalues $-j_i(j_i+1)$ and the corresponding eigenvectors $\mc{B}^{j_i}$.
\end{itemize}

After the diagonalization, we get a dimension 2 and dimension 4 degenerate eigenspaces for $j_{123}=1/2$ and $j_{123}=3/2$ respectively, which in eq.~\eqref{eq:basis_partial_wave} we have used the abstract index $a$ to label. Sometimes it is convenient to label the degeneracy by angular momenta of subgroup of particles, say $j_{12}$. Using the quantum numbers $(j_{123},j_{12})$, the 6 amplitudes can be further classified into 4 eigenspaces $(3/2,3/2)$, $(3/2,1/2)$, $(1/2,3/2)$, $(1/2,1/2)$, the first three being all one-dimensional
\eq{
	\mc{B}^{3/2,3/2}_{\psi^2\phi^3D^2} &= -6\mc{B}_1 + 2\mc{B}_2 + 2\mc{B}_3 + 9\mc{B}_4 + 3\mc{B}_5 + \mc{B}_6, \\
	\mc{B}^{3/2,1/2}_{\psi^2\phi^3D^2} &= -\mc{B}_2 + 2\mc{B}_3 + \mc{B}_6, \\
	\mc{B}^{1/2,3/2}_{\psi^2\phi^3D^2} &= 2\mc{B}_2 - \mc{B}_3 + 3\mc{B}_5 +\mc{B}_6.
}
In the $j_{123}=1/2$ dimension 4 eigenspace, we can also find sub-eigenspaces for $j_{13}$, where $j_{13}=3/2$ is also a one-dimensional eigenspace that we denote by a quantum number $\overline{3/2}$
\eq{
	\mc{B}^{1/2,\overline{3/2}}_{\psi^2\phi^3D^2} &= -\mc{B}_1 + 2\mc{B}_2 - \mc{B}_5,
}
which is independent of $\mc{B}^{1/2,3/2}$. Finally, the remaining dimension 2 linear space with $j_{123}=j_{12}=j_{13}=1/2$, we can find the two scalar angular momentum $j_{23}$ and obtain the eigenstates
\eq{
	\mc{B}^{1/2,1/2,0}_{\psi^2\phi^3D^2} &= \mc{B}_1 - \mc{B}_3 + \mc{B}_6, \\
	\mc{B}^{1/2,1/2,1}_{\psi^2\phi^3D^2} &= -\mc{B}_1 - \mc{B}_3 + \mc{B}_6.
}
The reason to obtain one-dimensional eigenspaces is to show the possibility of labelling all the degenerate states by angular momenta of subsets of particles. It also prevents mixing between the degenerate operators via renormalization, since they generate different partial wave states in the shared channel. For example, when consider the mixing between operator types $\psi_1\phi_2\phi_3\psi_4\phi_5D^2$ and $\psi_1 \phi_2\phi_3 \bar\psi_4 F_5 D$, we also diagonalize $\mc{W}$ for the latter and obtain the partial wave amplitude basis in the channels $\{1,2,3\}$ and $\{1,2\}$ the as following:
\eq{
\mc{B}^{3/2,3/2}_{F\psi\bar\psi\phi^2D} &=3\l15\r \l25\r[24]+\l15\r\l35\r[34], \\
\mc{B}^{3/2,1/2}_{F\psi\bar\psi\phi^2D} &=\l15\r\l35\r[35].
}
After this diagonalization, we conclude that only the pairs $\left(\mathcal{B}_{\psi^2\phi^3D^2}^{3/2,3/2},\mathcal{B}_{F\psi \bar \psi \phi^2D}^{3/2,3/2}\right)$ and $\left(\mathcal{B}_{\psi^2\phi^3D^2}^{3/2,1/2},\mathcal{B}_{F\psi \bar \psi \phi^2D}^{3/2,1/2}\right)$ can renormalize each other.

\end{document}